\documentclass[aps,prb,reprint,twocolumn,flushbottom,nobalancelastpage,noshowpacs,superscriptaddress]{revtex4-1}

\usepackage{amsmath,amssymb,amsfonts,fixmath}
\usepackage{bm,graphicx,color,afterpage}

\newcommand{\KILL}[1]{}

\newcommand{\TR}{\text{Tr}}
\newcommand{\RE}{\text{Re}}

\newcommand{\mydagger}{{\dagger}}
\newcommand{\phdagger}{{\phantom{\mydagger}\!}}
\newcommand{\bra}[1]{\langle{#1}|}
\newcommand{\ket}[1]{|{#1}\rangle}

\newcommand{\expval}[1]{\langle{#1}\rangle}
\newcommand{\bigval}[1]{\left\langle{#1}\right\rangle}
\newcommand{\expec}[2]{\langle{#1}\vert{#2}\vert{#1}\rangle}

\newcommand{\expGG}[1]{\expval{#1}_{\text{G}}^{\phantom{X}}}
\newcommand{\bigGG}[1]{\bigval{#1}_{\!\text{G}}^{\phantom{X}}}
\newcommand{\expGT}[1]{\expval{#1}_{\widetilde{\text{G}}}^{\phantom{X}}}
\newcommand{\ZGG}{Z_{{\text{G}}}^{\phantom{X}}}

\newcommand{\myhat}[1]{{\hat{#1}}}
\newcommand{\I}{{\myhat{\cal I}}}
\newcommand{\tildeI}{{\myhat{\widetilde{\cal I}}}}
\newcommand{\hatA}{\myhat{A}}
\newcommand{\hatH}{\myhat{H}}
\newcommand{\hatHdiag}{\myhat{H}_{\text{diag}}}
\newcommand{\hatS}{\myhat{S}}

\newcommand{\BG}{{\bm{G}}}
\newcommand{\BK}{{\bm{k}}}
\newcommand{\BN}{{\bm{n}}}
\newcommand{\BM}{{\bm{m}}}

\newcommand{\BP}{{\bm{p}}}

\newcommand{\tildeBN}{{\widetilde{\bm{n}}}}

\newcommand{\tildepsi}{{\widetilde{\psi}}}

\newcommand{\iniket}{\ket{\BP}}
\newcommand{\inibra}{\bra{\BP}}

\begin{document}

  \title{Generalized Gibbs ensemble prediction of prethermalization plateaus\\
    and their relation to nonthermal steady states in integrable systems}
  
  \author{Marcus Kollar}

  \author{F. Alexander Wolf}

  \altaffiliation[Current address: ]{Experimental Physics VI, Center for Electronic
    Correlations and Magnetism, Institute of Physics, University of
    Augsburg, 86135 Augsburg, Germany}

  \affiliation{Theoretical Physics III, Center for Electronic
    Correlations and Magnetism, Institute of Physics, University of
    Augsburg, 86135 Augsburg, Germany}

  \author{Martin Eckstein}

  \affiliation{Institute of Theoretical Physics, ETH Zurich,
    Wolfgang-Pauli-Str. 27, 8093 Zurich, Switzerland}

  \date{\today}

  \begin{abstract}
    A quantum many-body system which is prepared in the ground state
    of an integrable Hamiltonian does not directly thermalize after a
    sudden small parameter quench away from integrability. Rather, it
    will be trapped in a prethermalized state and can thermalize only
    at a later stage.  We discuss several examples for which this
    prethermalized state shares some properties with the nonthermal
    steady state that emerges in the corresponding integrable system.
    These examples support the notion that nonthermal steady states in
    integrable systems may be viewed as prethermalized states that
    never decay further.  Furthermore we show that prethermalization
    plateaus are under certain conditions correctly predicted by
    generalized Gibbs ensembles, which are the appropriate extension
    of standard statistical mechanics in the presence of many
    constants of motion.  This establishes that the relaxation
    behaviors of integrable and nearly integrable systems are
    continuously connected and described by the same statistical
    theory.
  \end{abstract}

  \maketitle

  \section{Introduction}

  Quantum statistical mechanics can successfully predict the
  equilibrium properties of a system with many degrees of freedom,
  based only on a few macroscopic parameters such as energy, volume,
  and particle number.  These predictions are obtained as averages
  over an ensemble of identical systems in which, according to the
  \emph{fundamental postulate of statistical mechanics}, each
  accessible microstate is equally probable. The ensemble is described
  by a statistical operator $\myhat{\rho}$ (with $\TR[\myhat{\rho}]$
  $=$ 1) which maximizes the entropy $S$ $=$
  $-\TR[\myhat{\rho}\ln\myhat{\rho}]$. In the microcanonical ensemble
  $\myhat{\rho}$ projects onto states with the correct macroscopic
  energy, but energy or other constants of motion can also be fixed
  only on average, as in the canonical or grand-canonical Gibbs
  ensemble.\cite{Jaynes1957b,*Jaynes1957c,Balian1991a} For macroscopic
  systems, the difference between the predictions of these standard
  ensembles is usually negligible, and they all describe the
  \emph{thermal state} of the system in equilibrium. The statistical
  prediction for the equilibrium expectation value of an observable
  $\hatA$ is then $\TR[\myhat{\rho}\hatA]$.

  An ensemble describes a superposition of quantum states with
  classical probabilities and hence is a mixed state for which
  $\TR[\myhat{\rho}^2]$ $<$ 1. Microscopically, however, a quantum
  system with Hamiltonian $\hatH(t)$ evolves according to the
  Schr{\"o}dinger equation, $i\hbar\frac{d}{dt}\ket{\psi(t)}$ $=$
  $\hatH(t)\ket{\psi(t)}$.  It is described by the density matrix
  $\myhat{\rho}(t)$ $=$ $\ket{\psi(t)}\bra{\psi(t)}$, i.e., a pure
  state with $\TR[\myhat{\rho}(t){}^2]$ $=$ 1.  This leads to the
  question how a disrupted quantum system can ever \emph{thermalize},
  i.e., relax to a new equilibrium state which is described by a
  thermal ensemble with $\TR[\myhat{\rho}{}^2]$ $<$ 1, although this
  quantity is constant during the unitary time evolution.  There are
  two principal physical resolutions to this apparent mathematical
  paradox: (i) If the system is in contact with a (typically much
  larger) environment and only observables of the system are of
  interest, then the environment degrees of freedom can be traced out
  from $\myhat{\rho}(t)$, leading to an effective statistical operator
  of the system that describes a mixed state.  (ii) If the system is
  isolated (as we assume here), then due to many-body interactions in
  the Hamiltonian the time evolution of $\ket{\psi(t)}$ can be
  sufficiently `ergodic' that for certain observables $\hatA$ the
  long-time limit of $\expval{\hatA}_t$ $=$
  $\bra{\psi(t)}\hatA\ket{\psi(t)}$ indeed tends to the statistical
  prediction $\TR[\myhat{\rho}\hatA]$. Several possibly related
  concepts were developed to understand this behavior: Inspired by von
  Neumann's quantum ergodic theorem, the theory of
  typicality\cite{vonNeumann1929a,*[English translation: ] %
    vonNeumann1929a_english,%
    Bocchieri1959a,Shankar1985a,Tasaki1998a,*Tasaki2010a,%
    Gemmer2010book,Goldstein2006a,*Goldstein2010Neumann,Popescu2007a,Reimann2007a,*Reimann2008a}
  puts bounds on the contributions to $\expval{\hatA}_t$ that are far
  from the thermal value.  The eigenstate thermalization
  hypothesis,\cite{Deutsch1991a,Srednicki1994a,Rigol2008a,Santos2010a,*Santos2010b,*Santos2010c}
  on the other hand, has relations to quantum chaos and posits that
  each eigenstate of $\hatH$ contributes to $\expval{\hatA}_t$ the
  microcanonical value \emph{at its eigenenergy}. Another useful point
  of view is that even in an isolated system a large part of it can
  act as an environment for the smaller
  remainder.\cite{Cramer2008a,Biroli2010a,Genway2010a,Cho2010a,Sciolla2010a,Haenggi2011a}
  Moreover, thermalization has been related to the many-body
  localization
  transition.\cite{Znidaric2008a,Pal2010a,SilvaMBL2010pre}

  Recent progress in the manipulation of cold atomic gases has made it
  possible to prepare quantum many-body systems in excellent isolation
  from the environment and to study their relaxation for a
  time-dependent Hamiltonian,\cite{Bloch2008a} thus providing a
  laboratory realization of the situation (ii) above. In particular,
  oscillations between Bose-condensed and Mott-insulating states after
  a steep sudden increase of the optical lattice
  depth~\cite{Greiner2002b} were observed.  In one-dimensional bosonic
  gases the dynamics leading to thermalization were measured for two
  coherently split gases\cite{Hofferberth2007a} and for a patterned
  initial state.\cite{Trotzky2011a} On the other hand, a nonthermal
  steady state was reached for a one-dimensional trap in which the
  system is close to an integrable point.\cite{Kinoshita2006} These
  developments have led to many theoretical studies regarding the
  relaxation of isolated quantum many-body systems (for recent
  reviews, see
  Refs.~\onlinecite{Dziarmaga2010a,Polkovnikov2010RMPpre,rigol2010review}).
  In the simplest setup, a quantum many-body system is studied after a
  sudden parameter change (``quench''). In this situation the time
  evolution for $t$ $\geq$ $0$ is governed by a time-independent
  Hamiltonian $\hatH$, but the initial state at $t$ $=$ $0$ is not an
  eigenstate of $\hatH$. Rather the system is typically prepared in
  the ground state or a thermal state of some other initial
  Hamiltonian $\hatH_0$.  Regarding the behavior of isolated
  interacting quantum systems after a global quench, three main cases
  can be distinguished: (a) Integrable systems which relax to a
  nonthermal steady state,\cite{Kinoshita2006,Dziarmaga2010a,%
    Polkovnikov2010RMPpre,rigol2010review,Girardeau1969,Sengupta2004,%
    Rigol2007,*Rigol2006,Cazalilla2006,*Iucci2009,Kollath2007,%
    Manmana2007,Gangardt2008a,Barthel2008a,Eckstein2008a,%
    Kollar2008a,Eckstein2008b,*Eckstein2008c,Lancaster2010a,Kennes2010a}
  which often can be described by generalized Gibbs ensembles (GGE)
  that take their large number of constants of motion into
  account;\cite{Jaynes1957b,*Jaynes1957c,Balian1991a,Rigol2007,*Rigol2006}
  (b) nearly integrable systems that do not thermalize directly, but
  instead are trapped in a prethermalized state on intermediate
  timescales, which can be predicted from perturbation
  theory;\cite{Berges2004a,Moeckel2008a,*Moeckel2009a,*Moeckel2010a,%
    Eckstein2009a,*Eckstein2010a,Sensarma2010a} and (c)
  nonintegrable systems which thermalize
  directly.\cite{Kollath2007,Rigol2008a,%
    Rigol2009t,*Rigol2010f,Eckstein2009a,*Eckstein2010a,Trotzky2011a}
  We review these three cases in Sec.~\ref{sec:integ-vs-therm}.

\begin{figure}[t]
    \centerline{\includegraphics[width=\columnwidth]{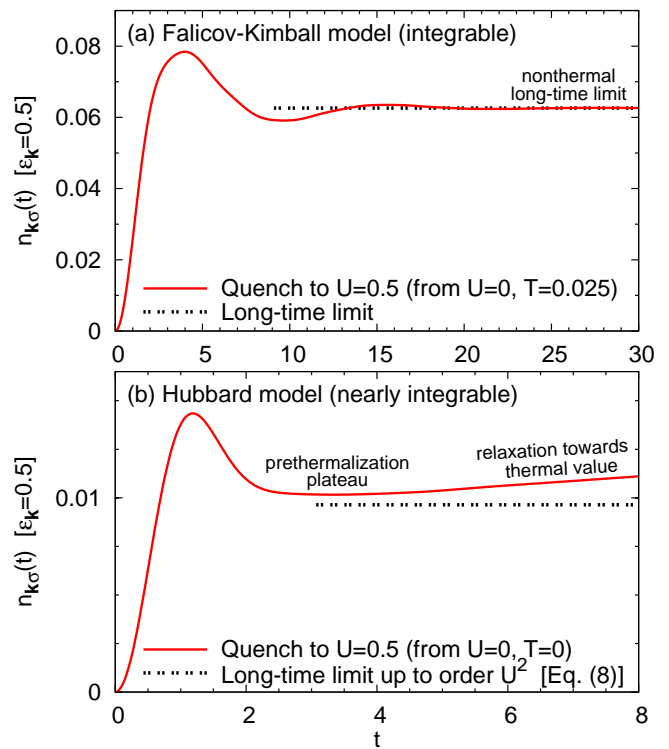}}
    \caption{Relaxation of the momentum occupation $n_{\BK\sigma}$
      after an interaction quench from $U$ $=$ $0$ to $U$ $=$ $0.5$ in
      (a) the Falicov-Kimball model\cite{Eckstein2008a} and (b)
      Hubbard model in iterated perturbation
      theory,\cite{Eckstein2009a,*Eckstein2010a} obtained in dynamical
      mean-field theory (DMFT) for a momentum $\BK$ which is outside
      the Fermi surface ($\epsilon_{\BK}=0.5$, half-filled band with
      semielliptic density of states, bandedges at $-2$ and $2$).  In
      the integrable Falicov-Kimball model a nonthermal long-time
      limit is observed, whereas in the nearly-integrable
      weak-coupling Hubbard model a prethermalization plateau occurs
      (which is predicted to good accuracy by second-order
      perturbation theory,\cite{Moeckel2008a,*Moeckel2009a} cf.\ %
      Sec.~\ref{subsec:prethermalization}), with subsequent relaxation
      towards the thermal value. For technical reasons the time
      evolution in (a) starts from a low-temperature thermal state.
      Further results for Falicov-Kimball and Hubbard models are
      discussed in Sec.~\ref{subsec:connection}.}
    \label{introfig}
  \end{figure}

  Fig.~\ref{introfig}
  shows two examples for the cases (a) and (b) for which the transient
  behavior is qualitatively rather similar. In particular, both the
  integrable and the nearly integrable system enter a long-lived
  nonthermal state. This leads us to the question whether and how the
  two cases are related and which properties they share.  Our main
  claim in this article is that (a) nonthermal steady states in
  integrable systems and (b) prethermalized states in nearly
  integrable systems are in precise correspondence, in the sense that
  both these nonthermal states are due to the existence of exact (in
  case (a)) or approximate (in case (b)) constants of motion (see
  Table~\ref{tab:quenches}). We support this claim by two types of
  evidence.  On the one hand (Sec.~\ref{subsec:connection}) we discuss
  several examples for which the predicted prethermalization plateau
  of an observable, \emph{when evaluated for an integrable system},
  yields precisely its nonthermal stationary value.  In other words,
  \emph{nonthermal steady states in integrable systems can be
    understood as prethermalized states that never decay}.  On the
  other hand (Sec.~\ref{subsec:GGEs}) we obtain perturbed constants of
  motion that are approximately conserved in a nearly integrable
  system, use them to construct the corresponding GGE, and show that
  it describes the prethermalization plateau for a certain class of
  observables.\cite{*[The statistical description of prethermalization
    plateaus with GGEs was already briefly reported in ] [] %
    epjst2010a} It follows that integrable and nearly integrable
  systems are connected in the sense that their relaxation dynamics
  involve long-lived nonthermal states that are described by the same
  statistical theory.
  \begin{table*}[t!]
    \caption{\label{tab:quenches}Nonthermal
      (quasi-)stationary states after a quench to an
      integrable or nearly integrable Hamiltonian~$\hatH$.}
    \begin{tabular*}{\hsize}{l@{~~}|@{~~}l@{~~}|@{~~}l}
      &
      Hamiltonian $\hatH$ after quench
      &
      (quasi-)stationary state
      \\
      \hline
      (a) integrable case
      &
      $\hatH$ integrable with exact constants of motion
      &
      nonthermal steady state in the
      \\
      &
      ~
      &
      long-time limit, $t$ $\to$ $\infty$
      \\
      \hline
      (b) nearly integrable case
      &
      $\hatH$ $=$ $\hatH_0$ $+$ $g\hatH_1$, $|g|$ $\ll$ $1$, $\hatH_0$ integrable,
      &
      prethermalized state for intermediate
      \\
      &
      $\hatH$ not integrable with approx. constants of motion
      &
      times $t$ with $|g|^{-1}$ $\ll$ $\text{const}\cdot t$ $\ll$ $g^{-2}$
      \\
      \hline
    \end{tabular*}
  \end{table*}

  \section{Integrability vs. Thermalization}\label{sec:integ-vs-therm}

  \subsection{Integrable systems: Nonthermal steady states}

  If $\hatH$ is integrable it has a large number of constants of
  motion, and the system then usually relaxes to a nonthermal steady
  state.\cite{Kinoshita2006,Dziarmaga2010a,Polkovnikov2010RMPpre,%
    rigol2010review,Girardeau1969,Sengupta2004,Rigol2007,*Rigol2006,%
    Cazalilla2006,*Iucci2009,Kollath2007,Manmana2007,Gangardt2008a,%
    Barthel2008a,Eckstein2008a,Kollar2008a,%
    Eckstein2008b,*Eckstein2008c,Lancaster2010a,Kennes2010a} This
  behavior is due to the fact that expectation values of all the
  constants of motion do not change with time. Therefore not all
  microstates in the relevant energy shell are in fact accessible, so
  that the above-mentioned fundamental postulate of statistical
  mechanics cannot be expected to give a reliable description of the
  steady state. In contrast to the classical case it is not obvious
  whether a given Hamiltonian is integrable, because any quantum
  Hamiltonian \emph{always} has as many constants of motion as the
  dimension of the Hilbert space, e.g., its powers, or the projectors
  onto its
  eigenstates.\cite{Weigert1992,Sutherland2004a,Manmana2007,Rigol2008a}
  Many solvable Hamiltonians $\hatH$, however, are integrable in a
  stronger sense, namely they can be mapped, $\hatH$ $\to$
  $\hatH_{\text{eff}}$, onto a effective Hamiltonian of the form
  \begin{align}
    \hatH_{\text{eff}}
    &=
    \sum_{\alpha=1}^L
    \epsilon_{\alpha}\I_{\alpha}
    \,,\label{Heff}
  \end{align}
  with ${[\I_{\alpha},\I_{\beta}]}$ $=$ $0$ for all $\alpha$ and
  $\beta$ and thus ${[\hatH,\I_{\alpha}]}$ $=$ $0$,
  where $L$  is proportional to the system size rather than the dimension
  of the Hilbert space of $\hatH_{\text{eff}}$. Typically 
   the constants of motion $\I_{\alpha}$ have integer
  eigenvalues that can be
  represented by fermionic or bosonic number operators, $\I_{\alpha}$
  $=$ $a^\mydagger_\alpha a^\phdagger_\alpha$. In these cases
  $\hatH_{\text{eff}}$ describes \emph{dressed} degrees of freedom that
  are noninteracting and have a simple time dependence. On the other
  hand, after transforming back the resulting time dependence of the
  original degrees of freedom is usually nontrivial.  

  Examples for models that can be solved on the Hamiltonian level as
  in Eq.~\eqref{Heff} include hard-core bosons in one dimension or XY
  spin chains, which can be mapped to noninteracting fermions by a
  Jordan-Wigner
  transformation,\cite{Girardeau1969,Sengupta2004,Rigol2007,Rigol2006,Barmettler2008a,LSM1961,Cazallila2011review}
  the Tomonaga-Luttinger model which corresponds to an effective
  free-boson Hamiltonian,\cite{Cazalilla2006,*Iucci2009,Haldane80} a
  one-dimensional electron-phonon model,\cite{Kennes2010a} and the
  $1/r$ Hubbard chain.\cite{Kollar2008a,Gebhard1992a,Gebhard1994a}
  The Falicov-Kimball
  model\cite{Eckstein2008a,Eckstein2008b,*Eckstein2008c,Freericks2003a}
  is also integrable in the sense that for a fixed equilibrium
  configuration of immobile particles the Hamiltonian is quadratic and
  can be diagonalized into the form~\eqref{Heff}.

  For effectively free Hamiltonians such as~(\ref{Heff}) a statistical
  prediction for the nonthermal steady state can be made with an
  appropriate GGE,\cite{Jaynes1957b,*Jaynes1957c,Balian1991a,Rigol2007,*Rigol2006}
  \begin{align}%
    \myhat{\rho}_{\text{G}}^\phdagger
    &=
    \frac{e^{-\sum_\alpha\lambda_\alpha\I_\alpha}}{Z_{\text{G}}^\phdagger}
    \,,
    &
    Z_{\text{G}}^\phdagger
    &=
    \TR[e^{-\sum_\alpha\lambda_\alpha\I_\alpha}]
    \,,\label{gge}%
  \end{align}%
  which maximizes the entropy with the constants of motion set to the
  correct average, $\expGG{\I_\alpha}$ $=$
  $\expval{\I_{\alpha}}_{t=0}$, by means of the Lagrange
  multipliers $\lambda_\alpha$.\cite{Balian1991a} The purpose of
  these additional constraints is to take into account (on average)
  that many microstates are inaccessible during the time evolution
  because they are incompatible with the values of the conserved
  quantities in the initial state.  GGEs correctly predict many (but
  not all) properties of nonthermal steady states in various
  integrable
  models.\cite{Dziarmaga2010a,Rigol2007,*Rigol2006,%
    Gangardt2008a,Barthel2008a,%
    Kollar2008a,Lancaster2010a,Kennes2010a}
  A microcanonical analogue of Eq.~\eqref{gge}, the so-called
  generalized microcanonical ensemble, was also
  studied.\cite{Cassidy2010a}
 
  \subsection{Nearly integrable systems: Prethermalization}\label{subsec:prethermalization}

  Now consider the case that the Hamiltonian $\hatH$ after the
  quench is not exactly integrable but close to an integrable point
  with Hamiltonian $\hatH_0$, i.e.,
  \begin{subequations}%
    \begin{align}%
      \hatH
      &= 
      \hatH_0
      +
      g\hatH_1
      \,,
      \label{H0+gH1}
      \\
      \hatH_0
      &=
      \sum_{\alpha=1}^L
      \epsilon_{\alpha}\I_{\alpha}
      \,,\label{H0}
    \end{align}%
    \label{H0+gH1-both}%
  \end{subequations}%
  with $|g|$ $\ll$ $1$, i.e., now the full Hamiltonian $\hatH$ is almost
  but not exactly of the form \eqref{Heff}. In this case the
  relaxation dynamics is nevertheless strongly influenced by the
  near-integrability, i.e., due to the presence of \emph{approximate
    constants of motion}, as discussed in more detail below. In such
  cases the system \emph{prethermalizes}, i.e., $\expval{\hatA}$
  relaxes first to a nonthermal quasistationary value
  $A_{\text{stat}}$ state that is increasingly long-lived as $\hatH$
  approaches the integrable point at $g$ $=$ $0$.  One of the
  characteristic features of prethermalization, known from field
  theory,\cite{Berges2004a} is that integrated quantities such as
  kinetic and potential energy attain their thermal values much
  earlier than individual occupation numbers. This phenomenon was
  recently studied in detail for Fermi liquids by Moeckel and
  Kehrein,\cite{Moeckel2008a,*Moeckel2009a,*Moeckel2010a} namely for interaction
  quenches from $U$ $=$ $0$ to small values of $U$ $>$ $0$ in the
  fermionic Hubbard model with Hamiltonian
  \begin{align}
    \hatH
    &=
    \sum_{ij\sigma}
    t_{ij}
    \myhat{c}_{i\sigma}^{\mydagger}
    \myhat{c}_{j\sigma}^{\phdagger}
    +
    U
    \sum_{i}
    \myhat{n}_{i\uparrow}
    \myhat{n}_{i\downarrow}
    \,, \label{hubbard}
  \end{align}
  which for $U$ $=$ $0$ reduces to an integrable
  Hamiltonian~\eqref{H0} in which the momentum occupation numbers
  $\myhat{n}_{\BK\sigma}$ $=$
  $\myhat{c}_{\BK\sigma}^{\mydagger}\myhat{c}_{\BK\sigma}^{\phdagger}$
  play the role of the conserved quantities $\I_\alpha$.

  It was stressed in
  Ref.~\onlinecite{Moeckel2008a,*Moeckel2009a,*Moeckel2010a} that in
  analogy to classical mechanics naive perturbation theory leads to
  secular terms that grow polynomially in time; instead instead one
  should use unitary perturbation theory, i.e., absorb the
  perturbation by a unitary transformation, perform the time
  evolution, and transform back. In Appendix~\ref{app:UPT} we derive a
  simple form of unitary perturbation theory (already used in
  Ref.~\onlinecite{Moeckel2009a}) for a nondegenerate Hamiltonian
  $\hatH_0$. If the time evolution is governed by the a nearly
  integrable Hamiltonian $\hatH$ [Eq.~\eqref{H0+gH1-both}], we obtain
  the expectation value of an observable $\hatA$ as (see
  Appendix~\ref{app:transients_general})
  \begin{align}%
    \expval{\hatA}_t
    &=
    \expval{\hatA}_0
    +
    4g^2
    \!\!
    \int\limits_{-\infty}^{\infty}\!\!d\omega\,
    \frac{\sin^2(\omega t/2)}{\omega^2}\,J(\omega)
    +O(g^3)
    \,,\label{eq:pretherm}
  \end{align}
  where the function $J(\omega)$ depends on the observable $A$ and the
  initial state $\ket{\psi(0)}$.  In the case that (i) $\hatA$ commutes
  with all constants of motion $\I_\alpha$ and (ii) the initial state
  $\ket{\psi(0)}$ is an eigenstate of $\hatH_0$, it can be written as
  \begin{align}
    J(\omega)
    &=
    \expval{
      \hatH_1
      \,
      (\hatA- \expval{\hatA}_0)
      \,
      \delta(\hatH_0-\expval{\hatH_0}-\omega)
      \,
      \hatH_1
    }_0
    \,.\label{eq:pretherm2}%
  \end{align}%
  These two assumptions (i) and (ii) are merely made to obtain the
  compact result~\eqref{eq:pretherm2}; it is straightforward to extend
  the analysis to any observable and any initial state. We note that
  an evaluation (see Appendix~\ref{app:transients_fermigas}) of
  $\expval{\myhat{n}_{\BK\sigma}}_t$ according to
  Eqs.~(\ref{eq:pretherm}-\ref{eq:pretherm2}) for quenches from $0$ to
  small $U$ in the fermionic Hubbard model~[Eq.~\eqref{hubbard}]
  recovers the result obtained with flow equations for continuous
  unitary transformations.\cite{Moeckel2008a,*Moeckel2009a,*Moeckel2010a} The
  prethermalization plateau, denoted by $A_\text{stat}$, can be
  obtained as the long-time average of Eq.~\eqref{eq:pretherm},
  $\lim_{t\to\infty}\int_{0}^{t}$ $dt'/t$ $\expval{A}_{t'}$, assuming
  that $|g|$ is so small that the scales $1/|g|$ and $1/g^2$ are well
  separated and the limit $t$ $\to$ $\infty$ is taken in the sense
  that $1/|g|$ $\ll$ $\text{const}\cdot t$ $\ll$
  $1/g^2$:\cite{Moeckel2009a}
  \begin{align}
    A_\text{stat}
    &=
    \expval{\hatA}_0
    +
    2g^2
    \!\!
    \int\limits_{-\infty}^{\infty}\!\!
    \frac{d\omega}{\omega^2}\,J(\omega)
    +O(g^3)
    \,.\label{eq:prethermplat}
    \intertext{If $\hatA$ commutes with all $\I_\alpha$ and $\ket{\psi(0)}$ is an
  eigenstate of $\hatH_0$, this expression simplifies to}
    A_\text{stat}
    &=
    2\expval{\hatA}_{\widetilde{0}}
    -
    \expval{\hatA}_0
    +O(g^3)
    \,,\label{theorem1}
  \end{align}
  where $\expval{\hatA}_{\widetilde{0}}$ $=$
  $\bra{\tildepsi{}(0)}A\ket{\tildepsi{}(0)}$ denotes the expectation
  value in the perturbative eigenstate $\ket{\tildepsi{}(0)}$ of
  $\hatH$ corresponding to the initial state
  $\ket{\psi(0)}$.\cite{Moeckel2009a}

  In general $A_\text{stat}$ differs from the thermal expectation
  value of $\hatA$ obtained with a microcanonical or canonical
  ensemble with the same average energy $E$ as the quenched system,
  i.e., $E$ $=$ $\bra{\psi(0)}\hatH\ket{\psi(0)}$ $=$
  $\bra{\psi(t)}\hatH\ket{\psi(t)}$.  Hence if subsequent
  thermalization occurs it is expected to be due to processes of order
  $g^3$ and higher and to happen at later times, $t$ $\gg$
  $1/g^2$.\cite{Moeckel2008a,*Moeckel2009a,%
    Eckstein2009a,Sensarma2010a,Olshanii2010a} The prethermalization
  plateau~(\ref{theorem1}) and also the predicted transient
  behavior~\eqref{eq:pretherm}\cite{Moeckel2008a,*Moeckel2009a} were
  confirmed for $\myhat{n}_{\BK\sigma}$ after interaction quenches in
  the Hubbard model in DMFT;\cite{Eckstein2009a,*Eckstein2010a}
  later-stage relaxation towards the thermal values was also observed
  (see also Fig.~\ref{introfig}b).
  
  \subsection{Nonintegrable systems: Thermalization}

  For nonintegrable systems thermalization is expected for
  sufficiently long times because only few relevant constants of
  motion exist, and was observed in several
  systems.\cite{Kollath2007,Rigol2008a,Rigol2009t,*Rigol2010f,%
    Eckstein2009a,*Eckstein2010a,Trotzky2011a} Due to limitations in
  simulation time and/or system size it is sometimes difficult to
  determine whether the required distance from an integrable point for
  which thermalization occurs is finite (as suggested, e.g., by the
  results of Refs.~\onlinecite{Kollath2007,Manmana2007,Roux2009b}) or
  infinitesimal in the thermodynamic limit (as suggested by a general
  analysis in Ref.~\onlinecite{Olshanii2010a}) This issue, as well as
  the mechanism for thermalization, is still being developed and
  debated.\cite{vonNeumann1929a,%
    Bocchieri1959a,Shankar1985a,Tasaki1998a,*Tasaki2010a,%
    Gemmer2010book,Goldstein2006a,*Goldstein2010Neumann,%
    Popescu2007a,Reimann2007a,*Reimann2008a,%
    Deutsch1991a,Srednicki1994a,%
    Rigol2008a,Santos2010a,*Santos2010b,*Santos2010c,%
    Cramer2008a,Biroli2010a,%
    Genway2010a,Cho2010a,Sciolla2010a,Haenggi2011a,%
    Znidaric2008a,Pal2010a,SilvaMBL2010pre,Danshita2010a} Interestingly, signatures
  of thermalization were also found for certain variables in
  integrable systems.\cite{Rossini2009a,*Rossini2010a}

  \section{Integrable vs.\ nearly integrable systems}\label{sec:integ-vs-noninteg}

  Our main claim in this article is the close correspondence between
  (a) nonthermal stationary values in integrable systems, i.e.,
  $\expval{\hatA}_\infty$ $=$ $\lim_{t\to\infty}$ $\expval{\hatA}_t$,
  and (b) prethermalization plateaus $A_\text{stat}$ in nearly
  integrable systems.  In Sec.~\ref{subsec:connection} we discuss
  several examples for which the predicted prethermalization plateau
  of an observable~\eqref{eq:prethermplat}, when evaluated for an
  integrable system of type~\eqref{Heff}, yields precisely its
  nonthermal stationary value.  We then obtain in
  Sec.~\ref{subsec:GGEs} that prethermalized states are described
  by an appropriate GGE built from \emph{approximate} constants of
  motion, analogous to nonthermal steady states in integrable systems
  that are described by a GGE built from \emph{exact} constants of
  motion.

  \subsection{Nonthermal steady states in integrable systems are
    prethermalized states that never decay}\label{subsec:connection}

  We now compare the two values $A_\text{stat}$
  [Eq.~\eqref{eq:prethermplat}] and $\expval{\hatA}_{t=\infty}$
  analytically or to high numerical accuracy for interaction quenches
  to weak and strong coupling in two Hubbard-type models, namely in
  the $1/r$ Hubbard chain\cite{Kollar2008a} and the Falicov-Kimball
  model in DMFT (i.e., in the limit of infinite spatial
  dimensions),\cite{Eckstein2008a,Eckstein2008b,Eckstein2008c} which
  are integrable in the sense of Eq.~(\ref{Heff}).  For both models
  the Hamiltonian is of the form~\eqref{hubbard} (however, for the
  Falicov-Kimball model the hopping amplitude is zero for one of the
  spin species).  As observable we consider the double occupation
  $\myhat{d}$ $=$
  $\expval{\sum_i\myhat{n}_{i\uparrow}\myhat{n}_{i\downarrow}}/L$
  ($L$: number of lattice sites). We obtain $d_\text{stat}$ from
  Eq.~\eqref{eq:prethermplat} for these two integrable systems, and
  show that it agrees with the nonthermal stationary value
  $\expval{\myhat{d}}_\infty$.

  \subsubsection{Weak coupling}

  We first consider an interaction quench from $0$ to small values of
  $U$. Then the prethermalization plateau of $\myhat{n}_{\BK\sigma}$
  is given by Eq.~\eqref{theorem1}, and $d_\text{stat}$ can be
  obtained using energy conservation after the quench. For the
  integrable $1/r$ Hubbard chain (with bandwidth $W$ and particle
  density $n$ $\leq$ $1$) we use known properties of the perturbed
  ground state $\ket{\tildepsi{}(0)}$ and obtain (see
  Appendix~\ref{app:gwfderiv})
  \begin{align}
    d_\text{stat}
    &=
    \frac{n^2}{4}
    -
    \frac{n^2(3-2n)U}{6W}
    +
    O(U^2)
    \,.\label{d_GR}    
  \end{align}
  When comparing this predicted prethermalization plateau with the
  exact long-time limit $\expval{\myhat{d}}_\infty$
  (Ref.~\onlinecite{Kollar2008a}) we see that both values \emph{agree
    to order $U$ for all densities $n$ $\leq$ $1$}. For this
  integrable system Eq.~\eqref{theorem1} thus predicts the nonthermal
  stationary value instead of a prethermalization plateau.

  \subsubsection{Strong coupling}

  For interaction quenches from $0$ to large values of $U$ the final
  Hamiltonian is also close to an integrable point, namely the atomic
  limit with conserved occupation numbers
  $\myhat{c}_{i\sigma}^{\mydagger}\myhat{c}_{i\sigma}^{\phdagger}$ on
  each lattice site. However, we consider an initial Hamiltonian other
  than the atomic limit, so that Eqs.~\eqref{eq:pretherm2}
  and~\eqref{theorem1} do not apply.  Instead, $d_\text{stat}$ is
  given by unitary strong-coupling perturbation
  theory\cite{Harris1967,Eckstein2009a} as
  \begin{align}%
    d_\text{stat}
    &=
    \expval{d}_0
    +
    \sum_{ij\sigma}
    \frac{t_{ij\sigma}}{UL}
    \expval{
      c_{i\sigma}^\mydagger
      c_{j\sigma}^\phdagger
      (\myhat{n}_{i\bar\sigma}-\myhat{n}_{j\bar\sigma})^2
    }_0
    +
    O(U^{-2})
    \,,\label{dstat}
  \end{align}%
  valid for an arbitrary initial state $\ket{\psi(0)}$.  We note that
  for a nonintegrable system $d_\text{stat}$ was observed as the
  center of collapse-and-revival oscillations that occur after
  interaction quenches to large $U$ in the Hubbard model in
  DMFT.\cite{Eckstein2009a}

  For quenches from $U$ $=$ $0$ to large $U$ in the integrable $1/r$
  Hubbard model Eq.~\eqref{dstat} predicts:\cite{Eckstein2009a}
  \begin{align}
    d_\text{stat}
    =
    \frac{n^2}{4}
    -
    \frac{2(3-2n)W}{3U}
    +O(U^{-2})
    \,.\label{d_GR_sc}
  \end{align}
  Comparing this prediction with the exact long-time limit
  $\expval{d}_\infty$ (Ref.~\onlinecite{Kollar2008a}) we find again
  that they are \emph{in agreement to order $U^{-1}$ for all densities
    $n$ $\leq$ $1$}.

\begin{figure}[t]
    \centerline{\includegraphics[width=\columnwidth]{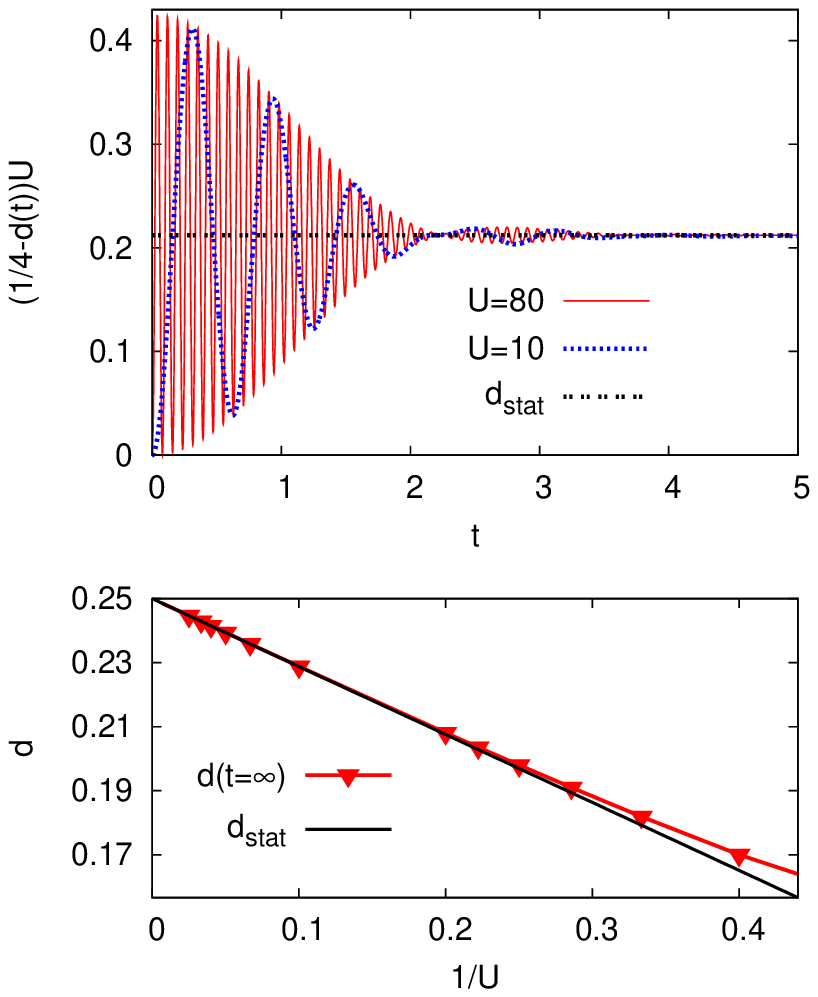}}
    \caption{Upper panel: Difference between the double occupation
      $\expval{d}_t$ and its initial value $\expval{d}_0$ $=$ $1/4$
      for quenches from the ground state ($U$ $=$ $0$) to $U$ $=$ $10$
      and $80$ in the Falicov-Kimball model in DMFT at half-filling,
      obtained from the exact solution for a semielliptic density of
      states with
      bandwidth~$4$~(Ref.~\onlinecite{Eckstein2008a,%
        Eckstein2008b,*Eckstein2008c}).
      For large $U$ the oscillations take place inside a common
      envelope function.\cite{Eckstein2009a} The horizontal line
      corresponds to the stationary value $d_\text{stat}$ to which
      $\expval{d}_t$ is predicted to relax according to the
      strong-coupling expansion~(\ref{dstat}).  Lower panel: The exact
      long-time limit $\expval{d}_t$ (triangle symbols) compared to
      the stationary value $d_\text{stat}$ of the strong-coupling
      expansion~(\ref{dstat}).}
    \label{fkm-fig-6}
  \end{figure}

  Finally, for the Falicov-Kimball model in DMFT with a semielliptic
  density of states, the value $d_\text{stat}$ predicted by
  Eq.~\eqref{dstat} is
  \begin{align}
    d_\text{stat}
    =
    \frac{n^2}{4}
    -
    \frac{(2-n)n}{2U}
    \expval{\hatH_0}_0
    +O(U^{-2})
    \,.\label{d_FK_sc}
  \end{align}
  Fig.~\ref{fkm-fig-6} shows the exact double occupation
  $\expval{d}_t$ for the Falicov-Kimball model in DMFT for quenches
  from $0$ to large $U$. In the long-time limit $\expval{d}_t$ tends
  precisely to the predicted value (\ref{d_FK_sc}) in the long-time
  limit large $U$.

  \subsubsection{Summary}

  For these three examples of integrable Hubbard-type systems we
  showed that prethermalized states, described by unitary perturbation
  theory for nearly integrable systems, also describes the nonthermal
  steady state in integrable systems.  This suggests the viewpoint
  that nonthermal steady states in integrable systems are simply
  prethermalized states that never decay.  In other words, the system
  appears to be trapped in essentially the same state both at and very
  close to an integrable point.  This suggests that the prethermalized
  state approaches the nonthermal steady state as one quenches closer
  and closer to the integrable point. We cannot show this continuity
  in general, but provide a continuous \emph{statistical} description
  of integrable and nonintegrable systems in the next subsection.


  \subsection{Construction of approximate constants of motion for
    nearly integrable systems and the corresponding generalized Gibbs
    ensemble}\label{subsec:GGEs}

  We now turn to the question whether for a small quench from an
  integrable point $\hatH_0$ to $\hatH$ $=$ $\hatH_0$ $+$ $g\hatH_1$
  (with $|g|$ $\ll$ $1$) the prethermalization
  plateau~\eqref{theorem1} is described by an appropriate Gibbs
  ensemble involving approximate constants of motion.  We use the
  eigenbasis $\ket{\bm{n}}$ of the constants of motion, i.e., $\bm{n}$
  $=$ $(n_1,n_2,\ldots,n_L)$, $\I_{\alpha}\ket{\bm{n}}$ $=$
  $n_{\alpha}\ket{\bm{n}}$, and assume that the energies
  $\epsilon_\alpha$ are incommensurate, so that the eigenenergies
  $E_{\BN}$ $=$ $\sum_\alpha\epsilon_\alpha n_\alpha$ of $\hatH_0$ are
  nondegenerate. This is not a strong restriction as the boundaries of
  the system can always be imagined to be so irregular as to lift all
  degeneracies.

  As described in Appendix~\ref{app:UPT} a unitary transformation
  $e^{\hatS}$ can be constructed which yields
  \begin{subequations}%
    \begin{align}%
      \hatH 
      &=
      \sum_{\alpha}\epsilon_{\alpha}
      \tildeI_{\alpha}
      \nonumber\\&\phantom{=}\;
      +
      \sum_{\tildeBN}
      \ket{\tildeBN} 
      (g E_{\BN}^{(1)} + g^2 E_{\BN}^{(2)})
      \bra{\tildeBN}
      +
      O(g^3)
      \,,\label{eq:H_in_tilde}
      \\
      \tildeI_{\alpha}
      &=
      e^{-\hatS} \I_{\alpha} e^{\hatS}
      \nonumber\\&
      =
      \I_{\alpha}
      -
      [\hatS,\I_{\alpha}]
      +
      [\hatS,[\hatS,\I_{\alpha}]]
      +
      O(g^3)    
      \,,
    \end{align}%
  \end{subequations}%
  where $\hatH\ket{\tildeBN}$ $=$ $\widetilde{E}_{\BN}\ket{\tildeBN}$,
  $\ket{\tildeBN}$ $=$ $e^{-\hatS}\ket{\BN}$, and $E_{\BN}^{(1,2)}$
  are the standard energy corrections in first and second order
  perturbation theory, recovering the perturbed
  Ray\-leigh-Schr{\"o}\-ding\-er energy eigenvalues,
  \begin{align}
    \widetilde{E}_{\BN}
    &=
    E_{\BN}
    +
    g
    E_{\BN}^{(1)}
    +
    g^2
    E_{\BN}^{(2)}
    +
    O(g^3)
    \,.\label{eq:En_tilde}
  \end{align}
  The structure of the transformed Hamiltonian is plausible: the
  first term on the left-hand side in Eq.~\eqref{eq:H_in_tilde}
  retains the additive `noninteracting' structure of the integrable
  Hamiltonian $\hatH_0$ with the same `one-particle' energies
  $\epsilon_\alpha$, whereas the perturbative energy corrections are
  not additive in this way but rather depend explicitly on the
  configuration of the state $e^{-\hatS}\ket{\BN}$.  Other perturbed
  Hamiltonians with a different structure were proposed in the
  literature, e.g., with modified energies
  $\widetilde{\epsilon}_\alpha$,\cite{Relano2010a} or perturbed
  energy eigenvalues $\widetilde{E}_{\BN}$ that remain additive in the
  quantum numbers $n_\alpha$.\cite{Heyl2010a}

  Since $[\tildeI_{\alpha},\tildeI_{\beta}]$ $=$
  $[\I_{\alpha},\I_{\beta}]$ $=$ 0 we have $[\hatH,\tildeI_{\alpha}]$
  $=$ $O(g^3)$, so that the $\tildeI_{\alpha}$ are the desired
  approximate constants of motion that indeed commute with $\hatH$ to
  order $g^2$. Note that in principle our canonical transformation can
  be continued to arbitrary high order in $g$, but an accurate
  description can nevertheless only be expected in a perturbative
  regime of sufficiently small $g$.  Next we construct the
  corresponding GGE with these perturbed constants of motion,
  \begin{align} 
    \myhat{\rho}_{\widetilde{\text{G}}}
    =
    \frac{1}{Z_{\widetilde{G}}}\,
    \exp\Big({-\sum_{\alpha}\lambda_\alpha \tildeI_{\alpha}}\Big)
    \,,\label{ggetilde}    
  \end{align}
  where the $\lambda_\alpha$ are fixed by the initial state according to
  \begin{align}
    \expval{\tildeI_{\alpha}}_{\widetilde{\text{G}}}  
    &=
    \TR[\myhat{\rho}_{\widetilde{\text{G}}}\tildeI_{\alpha}]
    \stackrel{!}{=} \langle \tildeI_{\alpha} \rangle_{0} 
    \,.\label{lambdafix}
  \end{align}
  Here we choose only the conserved quantities $\tildeI_{\alpha}$ that
  appear linearly and additively in the Hamiltonian
  \eqref{eq:H_in_tilde} to construct the GGE. Note that the
  Hamiltonian \eqref{eq:H_in_tilde} is not precisely of the form
  \eqref{Heff} but rather contains additional diagonal terms that
  involve the projectors $\ket{\tildeBN}\bra{\tildeBN}$.  These
  projectors are in general nonlinear in the $\tildeI_{\alpha}$ and
  are therefore not used in the GGE; the use of products of conserved
  quantities in the GGE is discussed in
  Refs.~\onlinecite{Gangardt2008a,Rigol2008a,Kollar2008a}, but not
  pursued here.
  
  We now come to the central point of this article: we compare the
  prethermalization plateau $A_{\text{stat}}$ [Eq.~\eqref{theorem1}]
  of an observable $\hatA$ (assumed to have the initial state as an
  eigenvector and to commute with all $\I_\alpha$) with the
  statistical prediction $\expval{\hatA}_{\widetilde{\text{G}}}$. We
  assume that the constants of motion $\I_{\alpha}$ can be represented
  by fermionic or bosonic number operators, $\I_{\alpha}$ $=$
  $a^\mydagger_\alpha a^\phdagger_\alpha$, and that the
  integrability-breaking term $\hatH_1$ can be expressed as a linear
  combination of products of these creation and annihilation
  operators. (Otherwise $\hatH_1$ would involve operators that act on
  other spaces than $\hatH_0$, so that $\hatH_0$ would have
  degeneracies contrary to our assumption.)

  For simplicity let us first consider an observable $\hatA$ $=$
  $\I_\alpha$, i.e., one of the conserved quantities of $\hatH_0$
  (e.g., a momentum occupation number $\myhat{n}_{\BK\sigma}$ in a
  Hubbard-type model).  Then we find (see Appendix~\ref{app:GGEproof})
  that indeed
  \begin{align}
    \I_{\alpha,\text{stat}}
    &=
    \expval{\I_{\alpha}}_{\tilde{\text{G}}}
    +
    O(g^3)
    \,.\label{ggesimplecase}
  \end{align}
  This shows that the prethermalization plateau of the conserved
  quantities of $\hatH_0$ (which are no longer conserved during the
  time evolution with $\hatH$ $=$ $\hatH_0$ $+$ $g\hatH_1$) is
  predicted correctly in order $g^2$ by the appropriate statistical
  theory [Eq.~\eqref{ggetilde}].  Hence on timescales $1/|g|$ $\ll$
  $\text{const}\cdot t$ $\ll$ $1/g^2$ the pure state $\ket{\psi(t)}$
  gives the same expectation values \emph{as a mixed state described
    by} $\myhat{\rho}_{\widetilde{\text{G}}}$.  For a more complicated
  observable,
  \begin{align}
    \hatA
    =
    \prod_{i=1}^{n}\I_{\alpha_i}
    \,,\label{eq:productA}
  \end{align} 
  we also find
  \begin{align}
    A_\text{stat}
    =
    \expval{\hatA}_{\tilde{\text{G}}}
    +
    O(g^3)
    \,,
  \end{align}
  provided the condition 
  \begin{align}
    \Big\langle
    \prod\limits_{i=1}^{n}
    \I_{\alpha_{i}}
    \Big\rangle_{\widetilde{0}}
    &=
    \prod\limits_{i=1}^{n}
    \langle
    \I_{\alpha_{i}}
    \rangle_{\widetilde{0}}
    +
    O(g^3)
    \label{condition}
  \end{align}
  is fulfilled. This is due to the fact that the GGE
  $\myhat{\rho}_{\widetilde{\text{G}}}$ is diagonal in the
  $\tildeI_{\alpha}$ and therefore cannot describe arbitrary
  correlations that are built up between two or more
  $\I_{\alpha_{i}}$, which is a well-known
  limitation.\cite{Gangardt2008a,Barthel2008a,Kollar2008a} At the
  integrable point ($g$ $=$ $0$ and $\ket{\psi(0)}$ $=$
  $\ket{\tildepsi{}(0)}$) the factorization
  condition~\eqref{condition} reduces to the condition derived in
  Ref.~\onlinecite{Kollar2008a} for the validity of a GGE~\eqref{gge}
  for an integrable Hamiltonian~\eqref{Heff}.

  The above assumption about the structure of $\hatH_1$ ensures that
  it does not contain operators that are absent in
  $\myhat{\rho}_{\widetilde{\text{G}}}$. Information about such
  operators would be missing from the GGE ensemble~\eqref{ggetilde},
  making their correct description unlikely.  However, this is not a
  strong restriction, as several coupled spaces can also be considered
  in a GGE (see, e.g., Ref.~\onlinecite{Kennes2010a}).

  We conclude that the phenomenon of prethermalization not only means
  that a long-lived nonthermal state is attained prior to possible
  thermalization at a later stage, but also that \emph{the properties
    of the prethermalized state are predicted correctly by an ensemble
    that is constructed according to the principles of statistical
    mechanics}.

  \section{Conclusion}

  We argued that integrable and nearly integrable systems are
  continuously connected in the following sense: (a) Integrable
  systems relax to nonthermal, but GGE-described stationary states;
  (b) Near-integrable systems are trapped in quasistationary states
  due to the perturbed constants of motion of the nearby integrable
  system, and can also be described by an appropriate perturbed GGE.
  Hence if one studies the relaxation of a nonintegrable system closer
  and closer to an integrable point, the prethermalization plateau
  will survive longer and longer and will approach the nonthermal
  long-time limit at the integrable point, with the appropriate GGE
  describing this steady state throughout.

  Previously GGEs were only used to describe integrable systems.  Here
  we showed that GGEs can make valid predictions also away from
  integrable points, at least perturbatively. In our opinion this
  illustrates the power of statistical mechanics, which makes correct
  predictions provided that the observables are not too complicated
  and only the accessible phase space is included in the statistical
  operator.

  \emph{Acknowledgements.} Insightful discussions with Stefan Kehrein,
  Michael Moeckel, Anatoli Polkovnikov, Marcos Rigol, Mark Srednicki,
  Michael Stark, Leticia Tarruell, Dieter Vollhardt, David Weiss, and
  Philipp Werner are gratefully acknowledged. This work was supported
  in part by DFG (SFB~484, TRR~80).

  \onecolumngrid~\twocolumngrid



%


  \onecolumngrid
  \section*{Appendix}
  \twocolumngrid

  \appendix

  \section{Unitary perturbation theory}
  \label{app:UPT}

  We use a canonical transformation $e^{\hatS}$, similar to
  Ref.~\onlinecite{Moeckel2009a}, which reproduces second-order
  Ray\-leigh-Schr{\"o}\-ding\-er perturbation theory at the operator level and
  thus enables us to construct the approximate constants of motion of
  $\hatH$ $=$ $\hatH_0$ $+$ $g\hatH_1$. We expand the antihermitian
  operator $\hatS$ in powers of $g$,
  \begin{align}
     \hatS
     &=
     g\hatS_1
     +
     \frac{1}{2}g^2\hatS_2
     +O(g^3)
     \,,
  \end{align}
  and apply the canonical transformation to $\hatH$,
  \begin{multline}
    e^{\hatS} \hatH e^{-\hatS}
    =
    \hatH_0
    +
    g
    \big(
    \hatH_1 + [\hatS_1, \hatH_0]
    \big)
    +
    g^2
    \bigg(
    \frac{1}{2}[\hatS_2,\hatH_0]
    \\
    +
    [\hatS_1, \hatH_1]
    +
    \frac{1}{2}[\hatS_1,[\hatS_1,\hatH_0]]
    \bigg)
    +
    O(g^3)
    \,.
  \end{multline}
  The transformed Hamiltonian shall still have all $\I_\alpha$ as
  constants of motion, i.e., we demand $[e^{\hatS} \hatH e^{-\hatS},
  \I_{\alpha}]$ $=$ $0$ for all $\alpha$, order by order. We use the
 basis $\I_{\alpha}\ket{\bm{n}}$ $=$
  $n_{\alpha}\ket{\bm{n}}$ and assume that the energies
  $\epsilon_\alpha$ are incommensurate, so that the eigenenergies
  $E_{\BN}$ $=$ $\sum_\alpha\epsilon_\alpha n_\alpha$ of $\hatH_0$ are
  nondegenerate.  To second order in $g$ we obtain for the transformed
  Hamiltonian and the unitary transformation
  \begin{align*}
    \hatHdiag
    &=
    e^{\hatS} \hatH e^{-\hatS}
    =   
    \hatH_0
    +
    g \hatHdiag^{(1)}
    +
    g^2 \hatHdiag^{(2)}
    +O(g^3)
    \,,
    \\
    \bra{\bm{n}}\hatS_1\ket{\BM}
    &=
    \begin{cases}
      \dfrac{\bra{\bm{n}}\hatH_1\ket{\BM}}{E_{\BN}-E_{\BM}} & \text{~if~}\bm{n}\neq\BM
      \\[1.5ex]
      0 & \text{~if~}\bm{n}=\BM
    \end{cases}
    \,,
    \\
    \bra{\bm{n}}\hatS_2\ket{\BM}
    &=
    \begin{cases}
      \dfrac{\bra{\bm{n}}[\hatS_1,\hatH_1+\hatHdiag^{(1)}]\ket{\BM}}{E_{\BN}-E_{\BM}}
      & \text{~if~}\bm{n}\neq\BM
      \\[1.5ex]
      0 & \text{~if~}\bm{n}=\BM
    \end{cases}
    \,,
    \\
    \hatHdiag^{(i)}
    &=
    \sum_{\BN} \ket{\bm{n}} 
    E_{\BN}^{(i)}
    \bra{\bm{n}}
    \,,
    \\
    E_{\BN}^{(1)} &= \bra{\bm{n}} \hatH_1 \ket{\bm{n}}
    \,,
    ~~
    E_{\BN}^{(2)}
    =
    \sum_{\BM(\neq \bm{n})}
    \frac{|\bra{\BM}\hatH_1\ket{\bm{n}}|^2}{E_{\BN}-E_{\BM}}
    \,,
  \end{align*}
  from which the eigenvalues $\widetilde{E}_{\BN}$
  [Eq.~\eqref{eq:En_tilde}] of the eigenstates $\ket{\tildeBN}$ $=$
  $e^{-\hatS}\ket{\BN}$ can be read off.

  \section{Transients in nearly integrable systems}
  \label{app:transients}

  \subsection{Derivation of Eqs.~\eqref{eq:pretherm}, \eqref{eq:pretherm2}, \eqref{theorem1}}
  \label{app:transients_general}

  Here we obtain the transient behavior in second order unitary
  perturbation theory, in close analogy to the derivation in
  Ref.~\onlinecite{Moeckel2009a}.  We assume that the initial state is
  an eigenstate of $\hatH_0$,
  \begin{align}
    \ket{\psi(0)}
    =
    \iniket
    \,,    
  \end{align}
  $\I_\alpha\iniket=p_\alpha\iniket$, and that the observable $\hatA$
  commutes with all constants of motion $\I_\alpha$. For now we set
  $\expval{\hatA}_0$ $=$ 0 and reinstate a possibly nonzero initial
  value at the end.  Inserting the unitary transformation for the
  Hamiltonian we obtain
  \begin{align}
    \expval{\hatA}_t
    &=
    \inibra{}
    e^{i\hatH t}
    \hatA
    e^{-i\hatH t}
    \iniket{}
    \nonumber\\
    &=
    \inibra{}
    e^{-\hatS}
    e^{i\hatHdiag t}
    e^{\hatS}
    \hatA
    e^{-\hatS}
    e^{-i\hatHdiag t}
    e^{\hatS}
    \iniket{}
    \nonumber\\
    &=
    \inibra{}
    e^{-\hatS}
    e^{\hatS(t)}
    \hatA
    e^{-\hatS(t)}
    e^{\hatS}
    \iniket{}
    \,,
  \end{align}
  with the abbreviation $\hatS(t)$ $=$ $e^{i\hatHdiag t} \hatS
  e^{-i\hatHdiag t}$.  Expanding the inner transformation as
  \begin{align}
    e^{\hatS(t)} \hatA e^{-\hatS(t)}
    =
    A
    +
    [S(t),A]
    +
    \frac12
    [S(t),[S(t),A]]
    +
    O(g^3)
  \end{align}
  and then similarly expanding the outer back transformation, we have
  \begin{align}
    \expval{\hatA}_t
    &=
    \inibra{}
    \hatA
    +[\hatS(t)-\hatS,\hatA]
    -\frac12[\hatS,[\hatS(t)-\hatS,\hatA]]
    \nonumber\\
    &~~~~~~~~~
    +\frac12[\hatS(t),[\hatS(t),\hatA]]
    \iniket{}
    +O(g^3)
    \nonumber\nonumber\\
    &=
    -\inibra{}
    (\hatS(t)-\hatS)\hatA(\hatS(t)-\hatS)
    \iniket{}
    +O(g^3)
    \nonumber\\
    &=
    -2\inibra{}\hatS\hatA\hatS\iniket{}
    +2\,\RE\inibra{}\hatS\hatA\hatS(t)\iniket{}
    +O(g^3)
    \,.\label{eq:transient1}
  \end{align}
  Here and below we frequently use that $\hatA$ annihilates
  $\iniket{}$, $\hatA$ commutes with $\hatHdiag$, and $\iniket{}$ is
  an eigenstate of $\hatHdiag$.  In the second term of the last
  equation we can rewrite
  \begin{align}
    &
    \inibra{}
    \hatS\hatA\hatS(t)
    \iniket{}
    \nonumber\\
    &=
    -
    \sum_{\bm{n}(\neq\BP)}
    |\inibra{}\hatS\ket{\bm{n}}|^2
    \bra{\bm{n}}\hatA\ket{\bm{n}}
    e^{-i(E_{\BP}-E_{\BN})t}
    \nonumber\\
    &=
    -
    \sum_{\bm{n}(\neq\BP)}
    \frac{|\inibra{}g\hatH_1\ket{\bm{n}}|^2}{(E_{\BP}-E_{\BN})^2}
    \bra{\bm{n}}\hatA\ket{\bm{n}}
    e^{-i(E_{\BP}-E_{\BN})t}
    +O(g^3)
    \nonumber\\
    &=
    -g^2
    \int\limits_{-\infty}^{\infty}
    \frac{d\omega}{\omega^2}
    \,
    J(\omega)
    \,
    e^{i\omega t}
    +O(g^3)
    \,,\label{eq:transient2}
  \end{align}
  Here we have defined 
  \begin{align}
    J(\omega)
    &=
    \sum_{\bm{n}(\neq\BP)}
    |\inibra{}\hatH_1\ket{\bm{n}}|^2
    \bra{\bm{n}}\hatA\ket{\bm{n}}
    \delta(\omega-(E_{\BN}-E_{\BP}))
    \nonumber\\
    &=
    \inibra{}
    \hatH_1
    \hatA
    \delta(\omega-(\hatH_0-\expval{\hatH_0}))
    \,
    \hatH_1
    \iniket{}
    \,,
  \end{align}
  as in Eq.~\eqref{eq:pretherm2}.  By setting $t$ $=$ $0$ in
  Eq.~\eqref{eq:transient2} we obtain a similar expression for the
  first term in Eq.~\eqref{eq:transient1}, which leads to
  Eq.~\eqref{theorem1}. Eq.~\eqref{eq:transient2} then also yields
  \begin{align}
    \expval{\hatA}_t
    &=
    g^2
    \int\limits_{-\infty}^{\infty}
    d\omega
    \,
    J(\omega)
    \,
    \frac{4\sin^2(\omega t/2)}{\omega^2}
    +O(g^3)
    \,,\label{eq:transient3}    
  \end{align}
  as in Eq.~\eqref{eq:pretherm}.

  \subsection{Evaluation for a small two-body interaction quench in a
    Fermi gas}
  \label{app:transients_fermigas}

  Here we evaluate the function $J(\omega)$ for a two-body interaction
  quench, i.e.,
  \begin{align}
    \hatH_0
    &=
    \sum_{\alpha}
    \epsilon_\alpha
    \,
    \myhat{c}_{\alpha}^{\mydagger}
    \myhat{c}_{\alpha}^{\phdagger}
    \,,
    &
    \hatH_1
    &=
    \sum_{\alpha\beta\gamma\delta}
    V_{\alpha\beta\gamma\delta}
    \,
    \myhat{c}_{\alpha}^{\mydagger}
    \myhat{c}_{\beta }^{\mydagger}
    \myhat{c}_{\gamma}^{\phdagger}
    \myhat{c}_{\delta}^{\phdagger}
    \,,
  \end{align}
  for fermionic operators,
  $\{\myhat{c}_{\alpha}^{\phdagger},\myhat{c}_{\beta}^{\mydagger}\}$
  $=$ $\delta_{\alpha\beta}$ and
  $\{\myhat{c}_{\alpha}^{\phdagger},\myhat{c}_{\beta}^{\phdagger}\}$
  $=$ $0$; hence $V_{\alpha\beta\gamma\delta}$ $=$
  $-V_{\beta\alpha\gamma\delta}$ $=$ $-V_{\alpha\beta\delta\gamma}$
  $=$ $V_{\beta\alpha\delta\gamma}$ and $V_{\alpha\beta\gamma\delta}$
  $=$ $=$ $(V_{\delta\gamma\beta\alpha})^*$.  The occupation numbers
  $\myhat{n}_\alpha$ $=$
  $\myhat{c}_{\alpha}^{\mydagger}\myhat{c}_{\alpha}^{\phdagger}$ (with
  eigenvalues 0, 1) play the role of constants of motion $\I_\alpha$
  of the unperturbed system (a Fermi gas) before the quench. As
  observable we choose the change in the occupation number  of a state $\mu$,
  \begin{align}
    A
    &=
    \I_\mu-\expval{\I_\mu}_0
    =
    \myhat{n}_{\mu}
    -p_\mu
    \,,
  \end{align}
  where $\ket{\psi(0)}$ $=$ $\iniket{}$ is the initial state with
  $\I_\alpha\iniket{}$ $=$ $p_\alpha\iniket{}$. 
  \begin{multline}
    J(\omega)
    =
    \sum_{\substack{
        \alpha\beta\gamma\delta\\
        \alpha'\beta'\gamma'\delta'}}
    V_{\alpha'\beta'\gamma'\delta'}
    V_{\alpha\beta\gamma\delta}
    \;
    \inibra{}
    \myhat{c}_{\alpha'}^{\mydagger}
    \myhat{c}_{\beta '}^{\mydagger}
    \myhat{c}_{\gamma'}^{\phdagger}
    \myhat{c}_{\delta'}^{\phdagger}
    \;\times
    \\
    (\myhat{n}_\mu-p_\mu)
    \,
    \delta(\omega-(\hatH_0-E_{\BP}))
    \,
    \myhat{c}_{\alpha}^{\mydagger}
    \myhat{c}_{\beta }^{\mydagger}
    \myhat{c}_{\gamma}^{\phdagger}
    \myhat{c}_{\delta}^{\phdagger}
    \iniket{}
    \,,
  \end{multline}
  where $\hatH_0-E_{\BP}$ inside the delta function evaluates to
  $\epsilon_\alpha$ $+$ $\epsilon_\beta$ $-$ $\epsilon_\gamma$ $-$
  $\epsilon_\delta$.  In the initial state the single-particle level
  $\mu$ may be occupied ($p_\mu$ $=$ 1) or unoccupied ($p_\mu$ $=$ 0),
  and inside the sum the operator $(\myhat{n}_\mu-p_\mu)$ must yield a
  nonzero contribution.  We consider first $p_\mu$ $=$ 1, in which
  case this requirement leads to a factor
  \begin{multline*}
    (\delta_{\gamma\mu}(1-\delta_{\delta\mu})+\delta_{\delta\mu}(1-\delta_{\gamma\mu}))
    (1-\delta_{\alpha\mu})(1-\delta_{\beta\mu})
    \\
    \times
    (1-\delta_{\gamma'\mu})(1-\delta_{\delta'\mu})
    (\delta_{\alpha'\mu}(1-\delta_{\beta'\mu})+\delta_{\beta'\mu}(1-\delta_{\alpha'\mu}))
    \,.
  \end{multline*}
  Using the symmetries of $V_{\alpha\beta\gamma\delta}$ we obtain the
  following contribution to $J(\omega)$,
  \begin{multline}
    -p_\mu
    \sum_{\substack{
        \alpha\beta\gamma\beta'\gamma'\delta'\\
        (\neq\mu)}}
    4V_{\mu\beta'\gamma'\delta'}
    V_{\alpha\beta\gamma\mu}
    \;
    \times\\
    \delta(\epsilon_\alpha+\epsilon_\beta-\epsilon_\gamma-\epsilon_\mu-\omega)
    \;
    \inibra{}
    \myhat{c}_{\beta '}^{\mydagger}
    \myhat{c}_{\gamma'}^{\phdagger}
    \myhat{c}_{\delta'}^{\phdagger}
    \myhat{c}_{\alpha }^{\mydagger}
    \myhat{c}_{\beta  }^{\mydagger}
    \myhat{c}_{\gamma }^{\phdagger}
    \iniket{}
    \,.\label{eq:pmu1}
  \end{multline}
  Next for $p_\mu$ $=$ 0 we find the factor
  \begin{multline*}
    (\delta_{\alpha\mu}(1-\delta_{\beta\mu})+\delta_{\beta\mu}(1-\delta_{\alpha\mu}))
    (1-\delta_{\gamma\mu})(1-\delta_{\delta\mu})
    \\
    \times
    (1-\delta_{\alpha'\mu})(1-\delta_{\beta'\mu})
    (\delta_{\gamma'\mu}(1-\delta_{\delta'\mu})+\delta_{\delta'\mu}(1-\delta_{\gamma'\mu}))
    \,,
  \end{multline*}
  so that the contribution  to  $J(\omega)$ is
  \begin{multline}
    (1-p_\mu)
    \sum_{\substack{
        \alpha\beta\gamma\beta'\gamma'\delta'\\
        (\neq\mu)}}
    4V_{\mu\beta\gamma\delta}
    V_{\alpha'\beta'\gamma'\mu}
    \;
    \times\\
    \delta(\epsilon_\mu+\epsilon_\beta-\epsilon_\gamma-\epsilon_\delta-\omega)
    \;
    \inibra{}
    \myhat{c}_{\alpha'}^{\mydagger}
    \myhat{c}_{\beta '}^{\mydagger}
    \myhat{c}_{\gamma'}^{\phdagger}
    \myhat{c}_{\beta  }^{\mydagger}
    \myhat{c}_{\gamma }^{\phdagger}
    \myhat{c}_{\delta }^{\phdagger}
    \iniket{}
    \,.\label{eq:pmu0}
  \end{multline}
  Evaluating the expectation values in Eqs.~\eqref{eq:pmu1}
  and~\eqref{eq:pmu0} in the product state $\iniket{}$ by contractions
  we finally obtain
  \begin{align}
    J(\omega)
    =
    ~~~~~
    &
    \nonumber\\
    -\;p_\mu
    \Bigg[
    &
    16
    \sum_\alpha
    (1-p_\alpha)
    \,
    |
    W_{\alpha\mu}
    |^2
    \,
    \delta(\epsilon_\alpha-\epsilon_\mu-\omega)
    \nonumber\\&
    +
    8
    \sum_{\alpha\beta\gamma}
    \big|
    V_{\alpha\beta\gamma\mu}
    \big|^2
    (1-p_\alpha)
    (1-p_\beta)
    p_\gamma
    \nonumber\\&
    ~~~~~~~~~
    \times\delta(\epsilon_\alpha+\epsilon_\beta-\epsilon_\gamma-\epsilon_\mu-\omega)
    \Bigg]
    \nonumber\\
    +
    \;
    (1-p_\mu)
    \Bigg[
    &
    16
    \sum_\alpha
    p_\alpha
    \,
    |W_{\alpha\mu}|^2
    \,
    \delta(\epsilon_\beta-\epsilon_\mu-\omega)
    \nonumber\\&
    +
    8
    \sum_{\alpha\beta\gamma}
    \big|
    V_{\alpha\beta\gamma\mu}
    \big|^2
    (1-p_\alpha)
    (1-p_\beta)
    p_\gamma
    \nonumber\\&
    ~~~~~~~~~
    \times\delta(\epsilon_\alpha+\epsilon_\beta-\epsilon_\gamma-\epsilon_\mu-\omega)
    \Bigg]
    \,,\label{eq:Jresult}
    \intertext{with the abbreviation}
    W_{\alpha\mu}
    &=
    \sum_\beta
    V_{\alpha\beta\beta\mu}
    \,
    p_\beta
    \,.
  \end{align}
  For completeness we now evaluate Eq.~\eqref{eq:Jresult} for the observable
  $\myhat{n}_{\BK\sigma}$ in the Hubbard model~\eqref{hubbard} by
  setting $\alpha$ $=$ $(\BK_1,\sigma_1)$ etc.,
  \begin{align}
    V_{\alpha\beta\gamma\delta}
    &=
    \frac{U}{4L}
    \Delta(\BK_1+\BK_2+\BK_3+\BK_4)
    \nonumber\\&\phantom{=}\;
    \times~
    \sum_\sigma
    \delta_{\sigma_1\sigma}
    \delta_{\sigma_2\bar{\sigma}}
    (
    \delta_{\sigma_3\sigma}
    \delta_{\sigma_4\bar{\sigma}}
    -
    \delta_{\sigma_3\bar{\sigma}}
    \delta_{\sigma_4\sigma}
    )
    \,,
  \end{align}
  so that in particular $V_{\alpha\beta\beta\delta}$ $=$ $0$,
  $W_{\alpha\mu}$ $=$ $0$. Here $\Delta(\BK)$ $=$ $\sum_\BG$
  $\delta_{\BK,\BG}$ is the von-Laue function involving reciprocal
  lattice vectors $\BG$. Eq.~\eqref{eq:Jresult} then takes the form
  \begin{align}
    J_{\BK\sigma}(\omega)
    &=
    -\frac{U^2}{L^2}
    \sum_{\substack{\BK_1\BK_2\BK_3\\\sigma_1\sigma_2\sigma_3}}
    \Delta(\BK_1+\BK_2-\BK_3-\BK)
    \nonumber\\&
    \times\delta(\epsilon_{\BK_1}+\epsilon_{\BK_2}-\epsilon_{\BK_3}-\epsilon_{\BK}-\omega)
    &\nonumber\\&\times~
    \Bigg[
    (1-p_{\BK_1\sigma_1})
    (1-p_{\BK_2\sigma_2})
    p_{\BK_3\sigma_3}
    p_{\BK\sigma}
    &\nonumber\\&~~~~~~
    -
    p_{\BK_1\sigma_1}
    p_{\BK_2\sigma_2}
    (1-p_{\BK_3\sigma_3})
    (1-p_{\BK\sigma})    
    \Bigg]
    \,,
  \end{align}
  where $p_{\BK\sigma}$ are the momentum occupation numbers in the
  initial state. When inserted into Eq.~\eqref{eq:pretherm} this leads
  to the same expression for the transient behavior that Moeckel and
  Kehrein\cite{Moeckel2008a,Moeckel2009a} obtained using continuous
  unitary transformations, but here we used only a single unitary
  transformation.

  \section{Properties of the weak-coupling ground state of the $1/r$ Hubbard chain}
  \label{app:gwfderiv}

  For the $1/r$ Hubbard chain the kinetic energy per lattice site
  $\epsilon_{\text{kin}}(U)$ can be obtained from
  the fact that the ground-state energy is given by the variational
  Gutzwiller energy up to $O(U^2)$,\cite{Gebhard1994a} which yields
  ($W$: bandwidth, $L$ $=$ number of lattice sites)
  \begin{align}
    \epsilon_{\text{kin}}(U)
    &=
    \frac{1}{L}
    \sum_{\BK\sigma}\epsilon_{\BK}\expval{\myhat{n}_{\BK\sigma}}_{\widetilde{0}}
    \label{gwfresult}
    \\
    &=
    -
    \frac{n(2-n)W}{4}
    -
    \frac{n^2(2n-3)U^2}{12W}
    +
    O\Big(\frac{U^3}{W^2}\Big)
    \,.\nonumber
  \end{align}
  For a quench from 0 to $U$ the prethermalization plateau of each
  momentum occupation number $\myhat{n}_{\BK\sigma}$ is given by
  Eq.~\eqref{theorem1}. Using the fact that the total energy is
  conserved after the quench, the prethermalization plateau of the
  double occupation $\myhat{d}$ is then given
  by
  \begin{align}%
    d_\text{stat}
    &=
    \expval{d}_0
    -\frac{2}{U}
    [\epsilon_{\text{kin}}(U)-\epsilon_{\text{kin}}(0)]
    +O(U^2)
    \,,\label{dgeneral}
  \end{align}%
  which, together with Eq.~\eqref{gwfresult}, yields Eq.~\eqref{d_GR}.

  \section{GGE prediction for prethermalization plateaus
    [Derivation of Eqs.~\eqref{ggesimplecase}, \eqref{condition}]}
  \label{app:GGEproof}

  In the following derivation of Eqs.~\eqref{ggesimplecase},
  \eqref{condition} we repeatedly use Eq.~(\ref{lambdafix}) which
  fixes the Lagrange multipliers. Several transformations between the
  eigenbases of the $\I_\alpha$ and the $\tildeI_\alpha$ are
  performed. We have
  \begin{align}
    \expGT{\hatA}
    &=
    \frac{
      \TR[\hatA\, e^{-\sum_{\alpha} \lambda_\alpha \tildeI_{\alpha}}]
    }{
      \TR[    e^{-\sum_{\alpha} \lambda_\alpha \tildeI_{\alpha}}]
    }
    \nonumber\\&
    =
    \frac{
      \TR[e^{\hatS} \hatA e^{-\hatS}\, e^{-\sum_{\alpha} \lambda_\alpha \I_{\alpha}}]
    }{
      \TR[               e^{-\sum_{\alpha} \lambda_\alpha \I_{\alpha}}]
    }
    =
    \expGG{e^{\hatS} \hatA e^{-\hatS}}
    \nonumber\\
    &=
    \expGG{\hatA+[\hatS,\hatA]+\tfrac{1}{2}[\hatS,[\hatS,\hatA]]} + O(g^3)
    \,,\label{AGGEtilde}
  \end{align}
  where $\expGG{\cdot}$ denotes the GGE expectation value~(\ref{gge})
  but with the $\lambda_\alpha$ still fixed by Eq.~(\ref{lambdafix}).
  We proceed to evaluate the three terms in $\expGT{\hatA}$ for an
  observable of the form~\eqref{eq:productA}. The first term can be
  rewritten as
  \begin{align}
    \expGG{\hatA}
    &=
    \bigGG{\prod\limits_{i=1}^{m} \I_{\alpha_{i}}}
    =
    \prod\limits_{i=1}^{m}
    \expGG{\I_{\alpha_{i}}}
    =
    \prod\limits_{i=1}^{m}
    \expGT{\tildeI_{\alpha_{i}}}
    \nonumber\\&
    =
    \prod\limits_{i=1}^{m} \expval{\tildeI_{\alpha_{i}}}_{0}
    = \prod\limits_{i=1}^{m} \expval{\I_{\alpha_{i}}}_{\widetilde{0}} + O(g^3)
    \,,
  \end{align}
  the second term vanishes, and the third term becomes
  \begin{align} 
    &\!\!\!\expGG{\tfrac{1}{2}[\hatS,[\hatS,\hatA]]}
    \nonumber\\
    &=\sum_{\BN}
    \frac{
      g^2\expec{\bm{n}}{\tfrac{1}{2}[\hatS_1,[\hatS_1,\hatA]]}
      +
      O(g^3)
    }{
      \ZGG
    }
    \,
    e^{-\sum_{{\alpha}}\lambda_{{\alpha}} n_{{\alpha}}}
    \nonumber\\&
    = g^2 F\big( \{ \expval{\I_\alpha}_{\text{G}} \} \big) + O(g^3) 
    \nonumber\\&
    = g^2 F\big( \{ \expval{\tildeI_{\alpha}}_{0} \} \big) + O(g^3)
    \nonumber\\&
    = g^2 F\big( \{ \expval{\I_{\alpha}}_{0} \} \big) + O(g^3)
    \nonumber\\&
    = g^2           \expval{\tfrac{1}{2}[\hatS_1,[\hatS_1,\hatA]]}_{0} + O(g^3 )
    \nonumber\\&
    = \expval{\tfrac{1}{2}[\hatS,[\hatS,\hatA]]}_{0} + O(g^3)
    \nonumber\\&
    = \expval{\hatA}_{\widetilde{0}} - \expval{\hatA}_0 +  O(g^3)
    \nonumber\\&
    =
    \expval{\prod\limits_{i=1}^{m} \!  \I_{\alpha_{i}}}_{\widetilde{0}}
    -
    \prod\limits_{i=1}^{m} \langle \I_{\alpha_{i}} \rangle_{0}
    +
    O(g^3)
  \end{align}
  In the second step we have used that $\hatH_1$ involves only the
  creation and annihilation operators that occur in $\hatH_0$ so that
  Wick's theorem can be applied, yielding some function $F$ of the
  occupation numbers, which are then related to initial-state
  expectation values in leading order in $g$. Then $F$ is eliminated
  by applying Wick's theorem backwards.  Finally, equating
  Eqs.~(\ref{theorem1}) and~(\ref{AGGEtilde}) yields the
  condition~(\ref{condition}).



\begin{thebibliography}{82}%
\makeatletter
\providecommand \@ifxundefined [1]{%
 \@ifx{#1\undefined}
}%
\providecommand \@ifnum [1]{%
 \ifnum #1\expandafter \@firstoftwo
 \else \expandafter \@secondoftwo
 \fi
}%
\providecommand \@ifx [1]{%
 \ifx #1\expandafter \@firstoftwo
 \else \expandafter \@secondoftwo
 \fi
}%
\providecommand \natexlab [1]{#1}%
\providecommand \enquote  [1]{``#1''}%
\providecommand \bibnamefont  [1]{#1}%
\providecommand \bibfnamefont [1]{#1}%
\providecommand \citenamefont [1]{#1}%
\providecommand \href@noop [0]{\@secondoftwo}%
\providecommand \href [0]{\begingroup \@sanitize@url \@href}%
\providecommand \@href[1]{\@@startlink{#1}\@@href}%
\providecommand \@@href[1]{\endgroup#1\@@endlink}%
\providecommand \@sanitize@url [0]{\catcode `\\12\catcode `\$12\catcode
  `\&12\catcode `\#12\catcode `\^12\catcode `\_12\catcode `\%12\relax}%
\providecommand \@@startlink[1]{}%
\providecommand \@@endlink[0]{}%
\providecommand \url  [0]{\begingroup\@sanitize@url \@url }%
\providecommand \@url [1]{\endgroup\@href {#1}{\urlprefix }}%
\providecommand \urlprefix  [0]{URL }%
\providecommand \Eprint [0]{\href }%
\providecommand \doibase [0]{http://dx.doi.org/}%
\providecommand \selectlanguage [0]{\@gobble}%
\providecommand \bibinfo  [0]{\@secondoftwo}%
\providecommand \bibfield  [0]{\@secondoftwo}%
\providecommand \translation [1]{[#1]}%
\providecommand \BibitemOpen [0]{}%
\providecommand \bibitemStop [0]{}%
\providecommand \bibitemNoStop [0]{.\EOS\space}%
\providecommand \EOS [0]{\spacefactor3000\relax}%
\providecommand \BibitemShut  [1]{\csname bibitem#1\endcsname}%
\let\auto@bib@innerbib\@empty
\bibitem [{\citenamefont {Jaynes}(1957{\natexlab{a}})}]{Jaynes1957b}%
  \BibitemOpen
  \bibfield  {author} {\bibinfo {author} {\bibfnamefont {E.}~\bibnamefont
  {Jaynes}},\ }\href {\doibase 10.1103/PhysRev.106.620} {\bibfield  {journal}
  {\bibinfo  {journal} {Phys. Rev.}\ }\textbf {\bibinfo {volume} {106}},\
  \bibinfo {pages} {620} (\bibinfo {year} {1957}{\natexlab{a}})}\BibitemShut
  {NoStop}%
\bibitem [{\citenamefont {Jaynes}(1957{\natexlab{b}})}]{Jaynes1957c}%
  \BibitemOpen
  \bibfield  {author} {\bibinfo {author} {\bibfnamefont {E.}~\bibnamefont
  {Jaynes}},\ }\href {\doibase 10.1103/PhysRev.108.171} {\bibfield  {journal}
  {\bibinfo  {journal} {Phys. Rev.}\ }\textbf {\bibinfo {volume} {108}},\
  \bibinfo {pages} {171} (\bibinfo {year} {1957}{\natexlab{b}})}\BibitemShut
  {NoStop}%
\bibitem [{\citenamefont {Balian}(1991)}]{Balian1991a}%
  \BibitemOpen
  \bibfield  {author} {\bibinfo {author} {\bibfnamefont {R.}~\bibnamefont
  {Balian}},\ }\href@noop {} {\emph {\bibinfo {title} {{From Microphysics to
  Macrophysics: Methods and Applications of Statistical Physics}}}},\
  Vol.~\bibinfo {volume} {1}\ (\bibinfo  {publisher} {Springer},\ \bibinfo
  {address} {Berlin},\ \bibinfo {year} {1991})\BibitemShut {NoStop}%
\bibitem [{\citenamefont {von Neumann}(1929)}]{vonNeumann1929a}%
  \BibitemOpen
  \bibfield  {author} {\bibinfo {author} {\bibfnamefont {J.}~\bibnamefont {von
  Neumann}},\ }\href@noop {} {\bibfield  {journal} {\bibinfo  {journal} {Z.
  Phys.}\ }\textbf {\bibinfo {volume} {57}},\ \bibinfo {pages} {30} (\bibinfo
  {year} {1929})}\BibitemShut {NoStop}%
\bibitem [{\citenamefont {von Neumann}(2010)}]{vonNeumann1929a_english}%
  \BibitemOpen
  \bibfield  {author} {\bibinfo {author} {\bibfnamefont {J.}~\bibnamefont {von
  Neumann}},\ }\href@noop {} {\bibfield  {journal} {\bibinfo  {journal} {Eur.
  Phys. J. H}\ }\textbf {\bibinfo {volume} {35}},\ \bibinfo {pages} {201}
  (\bibinfo {year} {2010})}\BibitemShut {NoStop}%
\bibitem [{\citenamefont {Bocchieri}\ and\ \citenamefont
  {Loinger}(1959)}]{Bocchieri1959a}%
  \BibitemOpen
  \bibfield  {author} {\bibinfo {author} {\bibfnamefont {P.}~\bibnamefont
  {Bocchieri}}\ and\ \bibinfo {author} {\bibfnamefont {A.}~\bibnamefont
  {Loinger}},\ }\href {\doibase 10.1103/PhysRev.114.948} {\bibfield  {journal}
  {\bibinfo  {journal} {Phys. Rev.}\ }\textbf {\bibinfo {volume} {114}},\
  \bibinfo {pages} {948} (\bibinfo {year} {1959})}\BibitemShut {NoStop}%
\bibitem [{\citenamefont {Jensen}\ and\ \citenamefont
  {Shankar}(1985)}]{Shankar1985a}%
  \BibitemOpen
  \bibfield  {author} {\bibinfo {author} {\bibfnamefont {R.~V.}\ \bibnamefont
  {Jensen}}\ and\ \bibinfo {author} {\bibfnamefont {R.}~\bibnamefont
  {Shankar}},\ }\href {\doibase 10.1103/PhysRevLett.54.1879} {\bibfield
  {journal} {\bibinfo  {journal} {Phys. Rev. Lett.}\ }\textbf {\bibinfo
  {volume} {54}},\ \bibinfo {pages} {1879} (\bibinfo {year}
  {1985})}\BibitemShut {NoStop}%
\bibitem [{\citenamefont {Tasaki}(1998)}]{Tasaki1998a}%
  \BibitemOpen
  \bibfield  {author} {\bibinfo {author} {\bibfnamefont {H.}~\bibnamefont
  {Tasaki}},\ }\href@noop {} {\bibfield  {journal} {\bibinfo  {journal} {Phys.
  Rev. Lett.}\ }\textbf {\bibinfo {volume} {80}},\ \bibinfo {pages} {1373}
  (\bibinfo {year} {1998})}\BibitemShut {NoStop}%
\bibitem [{\citenamefont {Tasaki}()}]{Tasaki2010a}%
  \BibitemOpen
  \bibfield  {author} {\bibinfo {author} {\bibfnamefont {H.}~\bibnamefont
  {Tasaki}},\ }\href@noop {} {}\Eprint
  {http://arxiv.org/abs/\linebreak[0]1003.5424} {arXiv:\linebreak[0]1003.5424}
  \BibitemShut {NoStop}%
\bibitem [{\citenamefont {Gemmer}, \citenamefont {Michel},\ and\ \citenamefont
  {Mahler}(2009)}]{Gemmer2010book}%
  \BibitemOpen
  \bibfield  {author} {\bibinfo {author} {\bibfnamefont {J.}~\bibnamefont
  {Gemmer}}, \bibinfo {author} {\bibfnamefont {M.}~\bibnamefont {Michel}}, \
  and\ \bibinfo {author} {\bibfnamefont {G.}~\bibnamefont {Mahler}},\
  }\href@noop {} {\emph {\bibinfo {title} {{Quantum Thermodynamics}}}},\
  \bibinfo {edition} {2nd}\ ed.,\ \bibinfo {series} {Lecture Notes in Physics},
  Vol.\ \bibinfo {volume} {784}\ (\bibinfo  {publisher} {Springer},\ \bibinfo
  {address} {Berlin},\ \bibinfo {year} {2009})\BibitemShut {NoStop}%
\bibitem [{\citenamefont {Goldstein}\ \emph {et~al.}(2006)\citenamefont
  {Goldstein}, \citenamefont {Lebowitz}, \citenamefont {Tumulka},\ and\
  \citenamefont {Zangh{\`i}}}]{Goldstein2006a}%
  \BibitemOpen
  \bibfield  {author} {\bibinfo {author} {\bibfnamefont {S.}~\bibnamefont
  {Goldstein}}, \bibinfo {author} {\bibfnamefont {J.~L.}\ \bibnamefont
  {Lebowitz}}, \bibinfo {author} {\bibfnamefont {R.}~\bibnamefont {Tumulka}}, \
  and\ \bibinfo {author} {\bibfnamefont {N.}~\bibnamefont {Zangh{\`i}}},\
  }\href@noop {} {\bibfield  {journal} {\bibinfo  {journal} {Phys. Rev. Lett.}\
  }\textbf {\bibinfo {volume} {96}},\ \bibinfo {pages} {050403} (\bibinfo
  {year} {2006})}\BibitemShut {NoStop}%
\bibitem [{\citenamefont {Goldstein}\ \emph {et~al.}(2010)\citenamefont
  {Goldstein}, \citenamefont {Lebowitz}, \citenamefont {Tumulka},\ and\
  \citenamefont {Zangh{\`i}}}]{Goldstein2010Neumann}%
  \BibitemOpen
  \bibfield  {author} {\bibinfo {author} {\bibfnamefont {S.}~\bibnamefont
  {Goldstein}}, \bibinfo {author} {\bibfnamefont {J.~L.}\ \bibnamefont
  {Lebowitz}}, \bibinfo {author} {\bibfnamefont {R.}~\bibnamefont {Tumulka}}, \
  and\ \bibinfo {author} {\bibfnamefont {N.}~\bibnamefont {Zangh{\`i}}},\
  }\href {\doibase 10.1140/epjh/e2010-00007-7} {\bibfield  {journal} {\bibinfo
  {journal} {Eur. Phys. J. H}\ }\textbf {\bibinfo {volume} {35}},\ \bibinfo
  {pages} {173} (\bibinfo {year} {2010})}\BibitemShut {NoStop}%
\bibitem [{\citenamefont {Popescu}, \citenamefont {Short},\ and\ \citenamefont
  {Winter}(2006)}]{Popescu2007a}%
  \BibitemOpen
  \bibfield  {author} {\bibinfo {author} {\bibfnamefont {S.}~\bibnamefont
  {Popescu}}, \bibinfo {author} {\bibfnamefont {A.~J.}\ \bibnamefont {Short}},
  \ and\ \bibinfo {author} {\bibfnamefont {A.}~\bibnamefont {Winter}},\ }\href
  {\doibase 10.1038/nphys444} {\bibfield  {journal} {\bibinfo  {journal}
  {Nature Physics}\ }\textbf {\bibinfo {volume} {2}},\ \bibinfo {pages} {754}
  (\bibinfo {year} {2006})}\BibitemShut {NoStop}%
\bibitem [{\citenamefont {Reimann}(2007)}]{Reimann2007a}%
  \BibitemOpen
  \bibfield  {author} {\bibinfo {author} {\bibfnamefont {P.}~\bibnamefont
  {Reimann}},\ }\href@noop {} {\bibfield  {journal} {\bibinfo  {journal} {Phys.
  Rev. Lett.}\ }\textbf {\bibinfo {volume} {99}},\ \bibinfo {pages} {160404}
  (\bibinfo {year} {2007})}\BibitemShut {NoStop}%
\bibitem [{\citenamefont {Reimann}(2008)}]{Reimann2008a}%
  \BibitemOpen
  \bibfield  {author} {\bibinfo {author} {\bibfnamefont {P.}~\bibnamefont
  {Reimann}},\ }\href@noop {} {\bibfield  {journal} {\bibinfo  {journal} {Phys.
  Rev. Lett.}\ }\textbf {\bibinfo {volume} {101}},\ \bibinfo {pages} {190403}
  (\bibinfo {year} {2008})}\BibitemShut {NoStop}%
\bibitem [{\citenamefont {Deutsch}(1991)}]{Deutsch1991a}%
  \BibitemOpen
  \bibfield  {author} {\bibinfo {author} {\bibfnamefont {J.~M.}\ \bibnamefont
  {Deutsch}},\ }\href {\doibase 10.1103/PhysRevA.43.2046} {\bibfield  {journal}
  {\bibinfo  {journal} {Phys. Rev. A}\ }\textbf {\bibinfo {volume} {43}},\
  \bibinfo {pages} {2046} (\bibinfo {year} {1991})}\BibitemShut {NoStop}%
\bibitem [{\citenamefont {Srednicki}(1994)}]{Srednicki1994a}%
  \BibitemOpen
  \bibfield  {author} {\bibinfo {author} {\bibfnamefont {M.}~\bibnamefont
  {Srednicki}},\ }\href {\doibase 10.1103/PhysRevE.50.888} {\bibfield
  {journal} {\bibinfo  {journal} {Phys. Rev. E}\ }\textbf {\bibinfo {volume}
  {50}},\ \bibinfo {pages} {888} (\bibinfo {year} {1994})}\BibitemShut
  {NoStop}%
\bibitem [{\citenamefont {Rigol}, \citenamefont {Dunjko},\ and\ \citenamefont
  {Olshanii}(2008)}]{Rigol2008a}%
  \BibitemOpen
  \bibfield  {author} {\bibinfo {author} {\bibfnamefont {M.}~\bibnamefont
  {Rigol}}, \bibinfo {author} {\bibfnamefont {V.}~\bibnamefont {Dunjko}}, \
  and\ \bibinfo {author} {\bibfnamefont {M.}~\bibnamefont {Olshanii}},\ }\href
  {\doibase 10.1038/nature06838} {\bibfield  {journal} {\bibinfo  {journal}
  {Nature}\ }\textbf {\bibinfo {volume} {452}},\ \bibinfo {pages} {854}
  (\bibinfo {year} {2008})}\BibitemShut {NoStop}%
\bibitem [{\citenamefont {Santos}\ and\ \citenamefont
  {Rigol}(2010{\natexlab{a}})}]{Santos2010a}%
  \BibitemOpen
  \bibfield  {author} {\bibinfo {author} {\bibfnamefont {L.~F.}\ \bibnamefont
  {Santos}}\ and\ \bibinfo {author} {\bibfnamefont {M.}~\bibnamefont {Rigol}},\
  }\href@noop {} {\bibfield  {journal} {\bibinfo  {journal} {Phys. Rev. E}\
  }\textbf {\bibinfo {volume} {81}},\ \bibinfo {pages} {036206} (\bibinfo
  {year} {2010}{\natexlab{a}})}\BibitemShut {NoStop}%
\bibitem [{\citenamefont {Rigol}\ and\ \citenamefont
  {Santos}(2010)}]{Santos2010b}%
  \BibitemOpen
  \bibfield  {author} {\bibinfo {author} {\bibfnamefont {M.}~\bibnamefont
  {Rigol}}\ and\ \bibinfo {author} {\bibfnamefont {L.~F.}\ \bibnamefont
  {Santos}},\ }\href@noop {} {\bibfield  {journal} {\bibinfo  {journal} {Phys.
  Rev. A}\ }\textbf {\bibinfo {volume} {82}},\ \bibinfo {pages} {011604}
  (\bibinfo {year} {2010})}\BibitemShut {NoStop}%
\bibitem [{\citenamefont {Santos}\ and\ \citenamefont
  {Rigol}(2010{\natexlab{b}})}]{Santos2010c}%
  \BibitemOpen
  \bibfield  {author} {\bibinfo {author} {\bibfnamefont {L.~F.}\ \bibnamefont
  {Santos}}\ and\ \bibinfo {author} {\bibfnamefont {M.}~\bibnamefont {Rigol}},\
  }\href@noop {} {\bibfield  {journal} {\bibinfo  {journal} {Phys. Rev. E}\
  }\textbf {\bibinfo {volume} {82}},\ \bibinfo {pages} {031130} (\bibinfo
  {year} {2010}{\natexlab{b}})}\BibitemShut {NoStop}%
\bibitem [{\citenamefont {Cramer}\ \emph {et~al.}(2008)\citenamefont {Cramer},
  \citenamefont {Dawson}, \citenamefont {Eisert},\ and\ \citenamefont
  {Osborne}}]{Cramer2008a}%
  \BibitemOpen
  \bibfield  {author} {\bibinfo {author} {\bibfnamefont {M.}~\bibnamefont
  {Cramer}}, \bibinfo {author} {\bibfnamefont {C.~M.}\ \bibnamefont {Dawson}},
  \bibinfo {author} {\bibfnamefont {J.}~\bibnamefont {Eisert}}, \ and\ \bibinfo
  {author} {\bibfnamefont {T.~J.}\ \bibnamefont {Osborne}},\ }\href {\doibase
  10.1103/PhysRevLett.100.030602} {\bibfield  {journal} {\bibinfo  {journal}
  {Phys. Rev. Lett.}\ }\textbf {\bibinfo {volume} {100}},\ \bibinfo {pages}
  {030602} (\bibinfo {year} {2008})}\BibitemShut {NoStop}%
\bibitem [{\citenamefont {Biroli}, \citenamefont {Kollath},\ and\ \citenamefont
  {Laeuchli}(2010)}]{Biroli2010a}%
  \BibitemOpen
  \bibfield  {author} {\bibinfo {author} {\bibfnamefont {G.}~\bibnamefont
  {Biroli}}, \bibinfo {author} {\bibfnamefont {C.}~\bibnamefont {Kollath}}, \
  and\ \bibinfo {author} {\bibfnamefont {A.}~\bibnamefont {Laeuchli}},\
  }\href@noop {} {\bibfield  {journal} {\bibinfo  {journal} {Phys. Rev. Lett.}\
  }\textbf {\bibinfo {volume} {105}},\ \bibinfo {pages} {250401} (\bibinfo
  {year} {2010})}\BibitemShut {NoStop}%
\bibitem [{\citenamefont {Genway}, \citenamefont {Ho},\ and\ \citenamefont
  {Lee}(2010)}]{Genway2010a}%
  \BibitemOpen
  \bibfield  {author} {\bibinfo {author} {\bibfnamefont {S.}~\bibnamefont
  {Genway}}, \bibinfo {author} {\bibfnamefont {A.~F.}\ \bibnamefont {Ho}}, \
  and\ \bibinfo {author} {\bibfnamefont {D.~K.~K.}\ \bibnamefont {Lee}},\
  }\href@noop {} {\bibfield  {journal} {\bibinfo  {journal} {Phys. Rev. Lett.}\
  }\textbf {\bibinfo {volume} {105}},\ \bibinfo {pages} {260402} (\bibinfo
  {year} {2010})}\BibitemShut {NoStop}%
\bibitem [{\citenamefont {Cho}\ and\ \citenamefont {Kim}(2010)}]{Cho2010a}%
  \BibitemOpen
  \bibfield  {author} {\bibinfo {author} {\bibfnamefont {J.}~\bibnamefont
  {Cho}}\ and\ \bibinfo {author} {\bibfnamefont {M.~S.}\ \bibnamefont {Kim}},\
  }\href@noop {} {\bibfield  {journal} {\bibinfo  {journal} {Phys. Rev. Lett.}\
  }\textbf {\bibinfo {volume} {104}},\ \bibinfo {pages} {170402} (\bibinfo
  {year} {2010})}\BibitemShut {NoStop}%
\bibitem [{\citenamefont {Sciolla}\ and\ \citenamefont
  {Biroli}(2010)}]{Sciolla2010a}%
  \BibitemOpen
  \bibfield  {author} {\bibinfo {author} {\bibfnamefont {B.}~\bibnamefont
  {Sciolla}}\ and\ \bibinfo {author} {\bibfnamefont {G.}~\bibnamefont
  {Biroli}},\ }\href@noop {} {\bibfield  {journal} {\bibinfo  {journal} {Phys.
  Rev. Lett.}\ }\textbf {\bibinfo {volume} {105}},\ \bibinfo {pages} {220401}
  (\bibinfo {year} {2010})}\BibitemShut {NoStop}%
\bibitem [{\citenamefont {Ponomarev}, \citenamefont {Denisov},\ and\
  \citenamefont {H{\"a}nggi}(2011)}]{Haenggi2011a}%
  \BibitemOpen
  \bibfield  {author} {\bibinfo {author} {\bibfnamefont {A.~V.}\ \bibnamefont
  {Ponomarev}}, \bibinfo {author} {\bibfnamefont {S.}~\bibnamefont {Denisov}},
  \ and\ \bibinfo {author} {\bibfnamefont {P.}~\bibnamefont {H{\"a}nggi}},\
  }\href@noop {} {\bibfield  {journal} {\bibinfo  {journal} {Phys. Rev. Lett.}\
  }\textbf {\bibinfo {volume} {106}},\ \bibinfo {pages} {010405} (\bibinfo
  {year} {2011})}\BibitemShut {NoStop}%
\bibitem [{\citenamefont {\v{Z}nidari\v{c}}, \citenamefont {Prosen},\ and\
  \citenamefont {Prelov\v{s}ek}(2008)}]{Znidaric2008a}%
  \BibitemOpen
  \bibfield  {author} {\bibinfo {author} {\bibfnamefont {M.}~\bibnamefont
  {\v{Z}nidari\v{c}}}, \bibinfo {author} {\bibfnamefont {T.}~\bibnamefont
  {Prosen}}, \ and\ \bibinfo {author} {\bibfnamefont {P.}~\bibnamefont
  {Prelov\v{s}ek}},\ }\href {\doibase 10.1103/PhysRevB.77.064426} {\bibfield
  {journal} {\bibinfo  {journal} {Phys. Rev. B}\ }\textbf {\bibinfo {volume}
  {77}},\ \bibinfo {pages} {064426} (\bibinfo {year} {2008})}\BibitemShut
  {NoStop}%
\bibitem [{\citenamefont {Pal}\ and\ \citenamefont {Huse}(2010)}]{Pal2010a}%
  \BibitemOpen
  \bibfield  {author} {\bibinfo {author} {\bibfnamefont {A.}~\bibnamefont
  {Pal}}\ and\ \bibinfo {author} {\bibfnamefont {D.~A.}\ \bibnamefont {Huse}},\
  }\href@noop {} {\bibfield  {journal} {\bibinfo  {journal} {Phys. Rev. B}\
  }\textbf {\bibinfo {volume} {82}},\ \bibinfo {pages} {174411} (\bibinfo
  {year} {2010})}\BibitemShut {NoStop}%
\bibitem [{\citenamefont {Canovi}\ \emph {et~al.}()\citenamefont {Canovi},
  \citenamefont {Rossini}, \citenamefont {Fazio}, \citenamefont {Santoro},\
  and\ \citenamefont {Silva}}]{SilvaMBL2010pre}%
  \BibitemOpen
  \bibfield  {author} {\bibinfo {author} {\bibfnamefont {E.}~\bibnamefont
  {Canovi}}, \bibinfo {author} {\bibfnamefont {D.}~\bibnamefont {Rossini}},
  \bibinfo {author} {\bibfnamefont {R.}~\bibnamefont {Fazio}}, \bibinfo
  {author} {\bibfnamefont {G.~E.}\ \bibnamefont {Santoro}}, \ and\ \bibinfo
  {author} {\bibfnamefont {A.}~\bibnamefont {Silva}},\ }\href@noop {} {}\Eprint
  {http://arxiv.org/abs/\linebreak[0]1006.1634} {arXiv:\linebreak[0]1006.1634}
  \BibitemShut {NoStop}%
\bibitem [{\citenamefont {Bloch}, \citenamefont {Dalibard},\ and\ \citenamefont
  {Zwerger}(2008)}]{Bloch2008a}%
  \BibitemOpen
  \bibfield  {author} {\bibinfo {author} {\bibfnamefont {I.}~\bibnamefont
  {Bloch}}, \bibinfo {author} {\bibfnamefont {J.}~\bibnamefont {Dalibard}}, \
  and\ \bibinfo {author} {\bibfnamefont {W.}~\bibnamefont {Zwerger}},\ }\href
  {\doibase 10.1103/RevModPhys.80.885} {\bibfield  {journal} {\bibinfo
  {journal} {Rev. Mod. Phys.}\ }\textbf {\bibinfo {volume} {80}},\ \bibinfo
  {pages} {885} (\bibinfo {year} {2008})}\BibitemShut {NoStop}%
\bibitem [{\citenamefont {Greiner}\ \emph {et~al.}(2002)\citenamefont
  {Greiner}, \citenamefont {Mandel}, \citenamefont {H{\"a}nsch},\ and\
  \citenamefont {Bloch}}]{Greiner2002b}%
  \BibitemOpen
  \bibfield  {author} {\bibinfo {author} {\bibfnamefont {M.}~\bibnamefont
  {Greiner}}, \bibinfo {author} {\bibfnamefont {O.}~\bibnamefont {Mandel}},
  \bibinfo {author} {\bibfnamefont {T.~W.}\ \bibnamefont {H{\"a}nsch}}, \ and\
  \bibinfo {author} {\bibfnamefont {I.}~\bibnamefont {Bloch}},\ }\href
  {\doibase 10.1038/nature00968} {\bibfield  {journal} {\bibinfo  {journal}
  {Nature}\ }\textbf {\bibinfo {volume} {419}},\ \bibinfo {pages} {51}
  (\bibinfo {year} {2002})}\BibitemShut {NoStop}%
\bibitem [{\citenamefont {Hofferberth}\ \emph {et~al.}(2007)\citenamefont
  {Hofferberth}, \citenamefont {Lesanovsky}, \citenamefont {Fischer},
  \citenamefont {Schumm},\ and\ \citenamefont
  {Schmiedmayer}}]{Hofferberth2007a}%
  \BibitemOpen
  \bibfield  {author} {\bibinfo {author} {\bibfnamefont {S.}~\bibnamefont
  {Hofferberth}}, \bibinfo {author} {\bibfnamefont {I.}~\bibnamefont
  {Lesanovsky}}, \bibinfo {author} {\bibfnamefont {B.}~\bibnamefont {Fischer}},
  \bibinfo {author} {\bibfnamefont {T.}~\bibnamefont {Schumm}}, \ and\ \bibinfo
  {author} {\bibfnamefont {J.}~\bibnamefont {Schmiedmayer}},\ }\href {\doibase
  10.1038/nature06149} {\bibfield  {journal} {\bibinfo  {journal} {Nature}\
  }\textbf {\bibinfo {volume} {449}},\ \bibinfo {pages} {324} (\bibinfo {year}
  {2007})}\BibitemShut {NoStop}%
\bibitem [{\citenamefont {Trotzky}\ \emph {et~al.}()\citenamefont {Trotzky},
  \citenamefont {Chen}, \citenamefont {Flesch}, \citenamefont {McCulloch},
  \citenamefont {Schollw{\"o}ck}, \citenamefont {Eisert},\ and\ \citenamefont
  {Bloch}}]{Trotzky2011a}%
  \BibitemOpen
  \bibfield  {author} {\bibinfo {author} {\bibfnamefont {S.}~\bibnamefont
  {Trotzky}}, \bibinfo {author} {\bibfnamefont {Y.-A.}\ \bibnamefont {Chen}},
  \bibinfo {author} {\bibfnamefont {A.}~\bibnamefont {Flesch}}, \bibinfo
  {author} {\bibfnamefont {I.~P.}\ \bibnamefont {McCulloch}}, \bibinfo {author}
  {\bibfnamefont {U.}~\bibnamefont {Schollw{\"o}ck}}, \bibinfo {author}
  {\bibfnamefont {J.}~\bibnamefont {Eisert}}, \ and\ \bibinfo {author}
  {\bibfnamefont {I.}~\bibnamefont {Bloch}},\ }\href@noop {} {}\Eprint
  {http://arxiv.org/abs/\linebreak[0]1101.2659} {arXiv:\linebreak[0]1101.2659}
  \BibitemShut {NoStop}%
\bibitem [{\citenamefont {Kinoshita}, \citenamefont {Wenger},\ and\
  \citenamefont {Weiss}(2006)}]{Kinoshita2006}%
  \BibitemOpen
  \bibfield  {author} {\bibinfo {author} {\bibfnamefont {T.}~\bibnamefont
  {Kinoshita}}, \bibinfo {author} {\bibfnamefont {T.}~\bibnamefont {Wenger}}, \
  and\ \bibinfo {author} {\bibfnamefont {D.~S.}\ \bibnamefont {Weiss}},\ }\href
  {\doibase doi:10.1038/nature04693} {\bibfield  {journal} {\bibinfo  {journal}
  {Nature}\ }\textbf {\bibinfo {volume} {440}},\ \bibinfo {pages} {900}
  (\bibinfo {year} {2006})}\BibitemShut {NoStop}%
\bibitem [{\citenamefont {Dziarmaga}(2010)}]{Dziarmaga2010a}%
  \BibitemOpen
  \bibfield  {author} {\bibinfo {author} {\bibfnamefont {J.}~\bibnamefont
  {Dziarmaga}},\ }\href {\doibase 10.1080/00018732.2010.514702} {\bibfield
  {journal} {\bibinfo  {journal} {Adv. Phys.}\ }\textbf {\bibinfo {volume}
  {59}},\ \bibinfo {pages} {1063} (\bibinfo {year} {2010})}\BibitemShut
  {NoStop}%
\bibitem [{\citenamefont {Polkovnikov}\ \emph {et~al.}()\citenamefont
  {Polkovnikov}, \citenamefont {Sengupta}, \citenamefont {Silva},\ and\
  \citenamefont {Vengalattore}}]{Polkovnikov2010RMPpre}%
  \BibitemOpen
  \bibfield  {author} {\bibinfo {author} {\bibfnamefont {A.}~\bibnamefont
  {Polkovnikov}}, \bibinfo {author} {\bibfnamefont {K.}~\bibnamefont
  {Sengupta}}, \bibinfo {author} {\bibfnamefont {A.}~\bibnamefont {Silva}}, \
  and\ \bibinfo {author} {\bibfnamefont {M.}~\bibnamefont {Vengalattore}},\
  }\href@noop {} {}\Eprint {http://arxiv.org/abs/\linebreak[0]1007.5331}
  {arXiv:\linebreak[0]1007.5331} \BibitemShut {NoStop}%
\bibitem [{\citenamefont {Rigol}()}]{rigol2010review}%
  \BibitemOpen
  \bibfield  {author} {\bibinfo {author} {\bibfnamefont {M.}~\bibnamefont
  {Rigol}},\ }\href@noop {} {}\Eprint
  {http://arxiv.org/abs/\linebreak[0]1008.1930} {arXiv:\linebreak[0]1008.1930}
  \BibitemShut {NoStop}%
\bibitem [{\citenamefont {Girardeau}(1969)}]{Girardeau1969}%
  \BibitemOpen
  \bibfield  {author} {\bibinfo {author} {\bibfnamefont {M.~D.}\ \bibnamefont
  {Girardeau}},\ }\href@noop {} {\bibfield  {journal} {\bibinfo  {journal}
  {Phys. Lett. A}\ }\textbf {\bibinfo {volume} {30}},\ \bibinfo {pages} {442}
  (\bibinfo {year} {1969})}\BibitemShut {NoStop}%
\bibitem [{\citenamefont {Sengupta}, \citenamefont {Powell},\ and\
  \citenamefont {Sachdev}(2004)}]{Sengupta2004}%
  \BibitemOpen
  \bibfield  {author} {\bibinfo {author} {\bibfnamefont {K.}~\bibnamefont
  {Sengupta}}, \bibinfo {author} {\bibfnamefont {S.}~\bibnamefont {Powell}}, \
  and\ \bibinfo {author} {\bibfnamefont {S.}~\bibnamefont {Sachdev}},\ }\href
  {\doibase 10.1103/PhysRevA.69.053616} {\bibfield  {journal} {\bibinfo
  {journal} {Phys. Rev. A}\ }\textbf {\bibinfo {volume} {69}},\ \bibinfo
  {pages} {053616} (\bibinfo {year} {2004})}\BibitemShut {NoStop}%
\bibitem [{\citenamefont {Rigol}\ \emph {et~al.}(2007)\citenamefont {Rigol},
  \citenamefont {Dunjko}, \citenamefont {Yurovsky},\ and\ \citenamefont
  {Olshanii}}]{Rigol2007}%
  \BibitemOpen
  \bibfield  {author} {\bibinfo {author} {\bibfnamefont {M.}~\bibnamefont
  {Rigol}}, \bibinfo {author} {\bibfnamefont {V.}~\bibnamefont {Dunjko}},
  \bibinfo {author} {\bibfnamefont {V.}~\bibnamefont {Yurovsky}}, \ and\
  \bibinfo {author} {\bibfnamefont {M.}~\bibnamefont {Olshanii}},\ }\href
  {\doibase 10.1103/PhysRevLett.98.050405} {\bibfield  {journal} {\bibinfo
  {journal} {Phys. Rev. Lett.}\ }\textbf {\bibinfo {volume} {98}},\ \bibinfo
  {pages} {050405} (\bibinfo {year} {2007})}\BibitemShut {NoStop}%
\bibitem [{\citenamefont {Rigol}, \citenamefont {Muramatsu},\ and\
  \citenamefont {Olshanii}(2006)}]{Rigol2006}%
  \BibitemOpen
  \bibfield  {author} {\bibinfo {author} {\bibfnamefont {M.}~\bibnamefont
  {Rigol}}, \bibinfo {author} {\bibfnamefont {A.}~\bibnamefont {Muramatsu}}, \
  and\ \bibinfo {author} {\bibfnamefont {M.}~\bibnamefont {Olshanii}},\ }\href
  {\doibase 10.1103/PhysRevA.74.053616} {\bibfield  {journal} {\bibinfo
  {journal} {Phys. Rev. A}\ }\textbf {\bibinfo {volume} {74}},\ \bibinfo
  {pages} {053616} (\bibinfo {year} {2006})}\BibitemShut {NoStop}%
\bibitem [{\citenamefont {Cazalilla}(2006)}]{Cazalilla2006}%
  \BibitemOpen
  \bibfield  {author} {\bibinfo {author} {\bibfnamefont {M.~A.}\ \bibnamefont
  {Cazalilla}},\ }\href {\doibase 10.1103/PhysRevLett.97.156403} {\bibfield
  {journal} {\bibinfo  {journal} {Phys. Rev. Lett.}\ }\textbf {\bibinfo
  {volume} {97}},\ \bibinfo {pages} {156403} (\bibinfo {year}
  {2006})}\BibitemShut {NoStop}%
\bibitem [{\citenamefont {Iucci}\ and\ \citenamefont
  {Cazalilla}(2009)}]{Iucci2009}%
  \BibitemOpen
  \bibfield  {author} {\bibinfo {author} {\bibfnamefont {A.}~\bibnamefont
  {Iucci}}\ and\ \bibinfo {author} {\bibfnamefont {M.~A.}\ \bibnamefont
  {Cazalilla}},\ }\href {\doibase 10.1103/PhysRevA.80.063619} {\bibfield
  {journal} {\bibinfo  {journal} {Phys. Rev. A}\ }\textbf {\bibinfo {volume}
  {80}},\ \bibinfo {pages} {063619} (\bibinfo {year} {2009})}\BibitemShut
  {NoStop}%
\bibitem [{\citenamefont {Kollath}, \citenamefont {L{\"a}uchli},\ and\
  \citenamefont {Altman}(2007)}]{Kollath2007}%
  \BibitemOpen
  \bibfield  {author} {\bibinfo {author} {\bibfnamefont {C.}~\bibnamefont
  {Kollath}}, \bibinfo {author} {\bibfnamefont {A.~M.}\ \bibnamefont
  {L{\"a}uchli}}, \ and\ \bibinfo {author} {\bibfnamefont {E.}~\bibnamefont
  {Altman}},\ }\href {\doibase 10.1103/PhysRevLett.98.180601} {\bibfield
  {journal} {\bibinfo  {journal} {Phys. Rev. Lett.}\ }\textbf {\bibinfo
  {volume} {98}},\ \bibinfo {pages} {180601} (\bibinfo {year}
  {2007})}\BibitemShut {NoStop}%
\bibitem [{\citenamefont {Manmana}\ \emph {et~al.}(2007)\citenamefont
  {Manmana}, \citenamefont {Wessel}, \citenamefont {Noack},\ and\ \citenamefont
  {Muramatsu}}]{Manmana2007}%
  \BibitemOpen
  \bibfield  {author} {\bibinfo {author} {\bibfnamefont {S.~R.}\ \bibnamefont
  {Manmana}}, \bibinfo {author} {\bibfnamefont {S.}~\bibnamefont {Wessel}},
  \bibinfo {author} {\bibfnamefont {R.~M.}\ \bibnamefont {Noack}}, \ and\
  \bibinfo {author} {\bibfnamefont {A.}~\bibnamefont {Muramatsu}},\ }\href
  {\doibase 10.1103/PhysRevLett.98.210405} {\bibfield  {journal} {\bibinfo
  {journal} {Phys. Rev. Lett.}\ }\textbf {\bibinfo {volume} {98}},\ \bibinfo
  {pages} {210405} (\bibinfo {year} {2007})}\BibitemShut {NoStop}%
\bibitem [{\citenamefont {Gangardt}\ and\ \citenamefont
  {Pustilnik}(2008)}]{Gangardt2008a}%
  \BibitemOpen
  \bibfield  {author} {\bibinfo {author} {\bibfnamefont {D.~M.}\ \bibnamefont
  {Gangardt}}\ and\ \bibinfo {author} {\bibfnamefont {M.}~\bibnamefont
  {Pustilnik}},\ }\href {\doibase 10.1103/PhysRevA.77.041604} {\bibfield
  {journal} {\bibinfo  {journal} {Phys. Rev. A}\ }\textbf {\bibinfo {volume}
  {77}},\ \bibinfo {pages} {041604} (\bibinfo {year} {2008})}\BibitemShut
  {NoStop}%
\bibitem [{\citenamefont {Barthel}\ and\ \citenamefont
  {Schollw{\"o}ck}(2008)}]{Barthel2008a}%
  \BibitemOpen
  \bibfield  {author} {\bibinfo {author} {\bibfnamefont {T.}~\bibnamefont
  {Barthel}}\ and\ \bibinfo {author} {\bibfnamefont {U.}~\bibnamefont
  {Schollw{\"o}ck}},\ }\href {\doibase 10.1103/PhysRevLett.100.100601}
  {\bibfield  {journal} {\bibinfo  {journal} {Phys. Rev. Lett.}\ }\textbf
  {\bibinfo {volume} {100}},\ \bibinfo {pages} {100601} (\bibinfo {year}
  {2008})}\BibitemShut {NoStop}%
\bibitem [{\citenamefont {Eckstein}\ and\ \citenamefont
  {Kollar}(2008{\natexlab{a}})}]{Eckstein2008a}%
  \BibitemOpen
  \bibfield  {author} {\bibinfo {author} {\bibfnamefont {M.}~\bibnamefont
  {Eckstein}}\ and\ \bibinfo {author} {\bibfnamefont {M.}~\bibnamefont
  {Kollar}},\ }\href {\doibase 10.1103/PhysRevLett.100.120404} {\bibfield
  {journal} {\bibinfo  {journal} {Phys. Rev. Lett.}\ }\textbf {\bibinfo
  {volume} {100}},\ \bibinfo {pages} {120404} (\bibinfo {year}
  {2008}{\natexlab{a}})}\BibitemShut {NoStop}%
\bibitem [{\citenamefont {Kollar}\ and\ \citenamefont
  {Eckstein}(2008)}]{Kollar2008a}%
  \BibitemOpen
  \bibfield  {author} {\bibinfo {author} {\bibfnamefont {M.}~\bibnamefont
  {Kollar}}\ and\ \bibinfo {author} {\bibfnamefont {M.}~\bibnamefont
  {Eckstein}},\ }\href {\doibase 10.1103/PhysRevA.78.013626} {\bibfield
  {journal} {\bibinfo  {journal} {Phys. Rev. A}\ }\textbf {\bibinfo {volume}
  {78}},\ \bibinfo {pages} {013626} (\bibinfo {year} {2008})}\BibitemShut
  {NoStop}%
\bibitem [{\citenamefont {Eckstein}\ and\ \citenamefont
  {Kollar}(2008{\natexlab{b}})}]{Eckstein2008b}%
  \BibitemOpen
  \bibfield  {author} {\bibinfo {author} {\bibfnamefont {M.}~\bibnamefont
  {Eckstein}}\ and\ \bibinfo {author} {\bibfnamefont {M.}~\bibnamefont
  {Kollar}},\ }\href {\doibase 10.1103/PhysRevB.78.205119} {\bibfield
  {journal} {\bibinfo  {journal} {Phys. Rev. B}\ }\textbf {\bibinfo {volume}
  {78}},\ \bibinfo {pages} {205119} (\bibinfo {year}
  {2008}{\natexlab{b}})}\BibitemShut {NoStop}%
\bibitem [{\citenamefont {Eckstein}\ and\ \citenamefont
  {Kollar}(2008{\natexlab{c}})}]{Eckstein2008c}%
  \BibitemOpen
  \bibfield  {author} {\bibinfo {author} {\bibfnamefont {M.}~\bibnamefont
  {Eckstein}}\ and\ \bibinfo {author} {\bibfnamefont {M.}~\bibnamefont
  {Kollar}},\ }\href {\doibase 10.1103/PhysRevB.78.245113} {\bibfield
  {journal} {\bibinfo  {journal} {Phys. Rev. B}\ }\textbf {\bibinfo {volume}
  {78}},\ \bibinfo {pages} {245113} (\bibinfo {year}
  {2008}{\natexlab{c}})}\BibitemShut {NoStop}%
\bibitem [{\citenamefont {Lancaster}\ and\ \citenamefont
  {Mitra}(2010)}]{Lancaster2010a}%
  \BibitemOpen
  \bibfield  {author} {\bibinfo {author} {\bibfnamefont {J.}~\bibnamefont
  {Lancaster}}\ and\ \bibinfo {author} {\bibfnamefont {A.}~\bibnamefont
  {Mitra}},\ }\href {\doibase 10.1103/PhysRevE.80.061134} {\bibfield  {journal}
  {\bibinfo  {journal} {Phys. Rev. E}\ }\textbf {\bibinfo {volume} {81}},\
  \bibinfo {pages} {061134} (\bibinfo {year} {2010})}\BibitemShut {NoStop}%
\bibitem [{\citenamefont {Kennes}\ and\ \citenamefont
  {Meden}(2010)}]{Kennes2010a}%
  \BibitemOpen
  \bibfield  {author} {\bibinfo {author} {\bibfnamefont {D.~M.}\ \bibnamefont
  {Kennes}}\ and\ \bibinfo {author} {\bibfnamefont {V.}~\bibnamefont {Meden}},\
  }\href {\doibase 10.1103/PhysRevB.82.085109} {\bibfield  {journal} {\bibinfo
  {journal} {Phys. Rev. B}\ }\textbf {\bibinfo {volume} {82}},\ \bibinfo
  {pages} {2010} (\bibinfo {year} {2010})}\BibitemShut {NoStop}%
\bibitem [{\citenamefont {Berges}, \citenamefont {Bors{\'a}nyi},\ and\
  \citenamefont {Wetterich}(2004)}]{Berges2004a}%
  \BibitemOpen
  \bibfield  {author} {\bibinfo {author} {\bibfnamefont {J.}~\bibnamefont
  {Berges}}, \bibinfo {author} {\bibfnamefont {S.}~\bibnamefont
  {Bors{\'a}nyi}}, \ and\ \bibinfo {author} {\bibfnamefont {C.}~\bibnamefont
  {Wetterich}},\ }\href {\doibase 10.1103/PhysRevLett.93.142002} {\bibfield
  {journal} {\bibinfo  {journal} {Phys. Rev. Lett.}\ }\textbf {\bibinfo
  {volume} {93}},\ \bibinfo {pages} {142002} (\bibinfo {year}
  {2004})}\BibitemShut {NoStop}%
\bibitem [{\citenamefont {Moeckel}\ and\ \citenamefont
  {Kehrein}(2008)}]{Moeckel2008a}%
  \BibitemOpen
  \bibfield  {author} {\bibinfo {author} {\bibfnamefont {M.}~\bibnamefont
  {Moeckel}}\ and\ \bibinfo {author} {\bibfnamefont {S.}~\bibnamefont
  {Kehrein}},\ }\href {\doibase 10.1103/PhysRevLett.100.175702} {\bibfield
  {journal} {\bibinfo  {journal} {Phys. Rev. Lett.}\ }\textbf {\bibinfo
  {volume} {100}},\ \bibinfo {pages} {175702} (\bibinfo {year}
  {2008})}\BibitemShut {NoStop}%
\bibitem [{\citenamefont {Moeckel}\ and\ \citenamefont
  {Kehrein}(2009)}]{Moeckel2009a}%
  \BibitemOpen
  \bibfield  {author} {\bibinfo {author} {\bibfnamefont {M.}~\bibnamefont
  {Moeckel}}\ and\ \bibinfo {author} {\bibfnamefont {S.}~\bibnamefont
  {Kehrein}},\ }\href {\doibase 10.1016/j.aop.2009.03.009} {\bibfield
  {journal} {\bibinfo  {journal} {Ann. Phys.}\ }\textbf {\bibinfo {volume}
  {324}},\ \bibinfo {pages} {2146} (\bibinfo {year} {2009})}\BibitemShut
  {NoStop}%
\bibitem [{\citenamefont {Moeckel}\ and\ \citenamefont
  {Kehrein}(2010)}]{Moeckel2010a}%
  \BibitemOpen
  \bibfield  {author} {\bibinfo {author} {\bibfnamefont {M.}~\bibnamefont
  {Moeckel}}\ and\ \bibinfo {author} {\bibfnamefont {S.}~\bibnamefont
  {Kehrein}},\ }\href@noop {} {\bibfield  {journal} {\bibinfo  {journal} {New
  J. Phys.}\ }\textbf {\bibinfo {volume} {12}},\ \bibinfo {pages} {055016}
  (\bibinfo {year} {2010})}\BibitemShut {NoStop}%
\bibitem [{\citenamefont {Eckstein}, \citenamefont {Kollar},\ and\
  \citenamefont {Werner}(2009)}]{Eckstein2009a}%
  \BibitemOpen
  \bibfield  {author} {\bibinfo {author} {\bibfnamefont {M.}~\bibnamefont
  {Eckstein}}, \bibinfo {author} {\bibfnamefont {M.}~\bibnamefont {Kollar}}, \
  and\ \bibinfo {author} {\bibfnamefont {P.}~\bibnamefont {Werner}},\
  }\href@noop {} {\bibfield  {journal} {\bibinfo  {journal} {Phys. Rev. Lett.}\
  }\textbf {\bibinfo {volume} {103}},\ \bibinfo {pages} {056403} (\bibinfo
  {year} {2009})}\BibitemShut {NoStop}%
\bibitem [{\citenamefont {Eckstein}, \citenamefont {Kollar},\ and\
  \citenamefont {Werner}(2010)}]{Eckstein2010a}%
  \BibitemOpen
  \bibfield  {author} {\bibinfo {author} {\bibfnamefont {M.}~\bibnamefont
  {Eckstein}}, \bibinfo {author} {\bibfnamefont {M.}~\bibnamefont {Kollar}}, \
  and\ \bibinfo {author} {\bibfnamefont {P.}~\bibnamefont {Werner}},\
  }\href@noop {} {\bibfield  {journal} {\bibinfo  {journal} {Phys. Rev. B}\
  }\textbf {\bibinfo {volume} {81}},\ \bibinfo {pages} {115131} (\bibinfo
  {year} {2010})}\BibitemShut {NoStop}%
\bibitem [{\citenamefont {Sensarma}\ \emph {et~al.}(2010)\citenamefont
  {Sensarma}, \citenamefont {Pekker}, \citenamefont {Altman}, \citenamefont
  {Demler}, \citenamefont {Strohmaier}, \citenamefont {Greif}, \citenamefont
  {J{\"o}rdens}, \citenamefont {Tarruell}, \citenamefont {Moritz},\ and\
  \citenamefont {Esslinger}}]{Sensarma2010a}%
  \BibitemOpen
  \bibfield  {author} {\bibinfo {author} {\bibfnamefont {R.}~\bibnamefont
  {Sensarma}}, \bibinfo {author} {\bibfnamefont {D.}~\bibnamefont {Pekker}},
  \bibinfo {author} {\bibfnamefont {E.}~\bibnamefont {Altman}}, \bibinfo
  {author} {\bibfnamefont {E.}~\bibnamefont {Demler}}, \bibinfo {author}
  {\bibfnamefont {N.}~\bibnamefont {Strohmaier}}, \bibinfo {author}
  {\bibfnamefont {D.}~\bibnamefont {Greif}}, \bibinfo {author} {\bibfnamefont
  {R.}~\bibnamefont {J{\"o}rdens}}, \bibinfo {author} {\bibfnamefont
  {L.}~\bibnamefont {Tarruell}}, \bibinfo {author} {\bibfnamefont
  {H.}~\bibnamefont {Moritz}}, \ and\ \bibinfo {author} {\bibfnamefont
  {T.}~\bibnamefont {Esslinger}},\ }\href {\doibase 10.1103/PhysRevB.82.224302}
  {\bibfield  {journal} {\bibinfo  {journal} {Phys. Rev. B}\ }\textbf {\bibinfo
  {volume} {82}},\ \bibinfo {pages} {224302} (\bibinfo {year}
  {2010})}\BibitemShut {NoStop}%
\bibitem [{\citenamefont {Rigol}(2009{\natexlab{a}})}]{Rigol2009t}%
  \BibitemOpen
  \bibfield  {author} {\bibinfo {author} {\bibfnamefont {M.}~\bibnamefont
  {Rigol}},\ }\href@noop {} {\bibfield  {journal} {\bibinfo  {journal} {Phys.
  Rev. Lett.}\ }\textbf {\bibinfo {volume} {103}},\ \bibinfo {pages} {100403}
  (\bibinfo {year} {2009}{\natexlab{a}})}\BibitemShut {NoStop}%
\bibitem [{\citenamefont {Rigol}(2009{\natexlab{b}})}]{Rigol2010f}%
  \BibitemOpen
  \bibfield  {author} {\bibinfo {author} {\bibfnamefont {M.}~\bibnamefont
  {Rigol}},\ }\href@noop {} {\bibfield  {journal} {\bibinfo  {journal} {Phys.
  Rev. A}\ }\textbf {\bibinfo {volume} {80}},\ \bibinfo {pages} {053607}
  (\bibinfo {year} {2009}{\natexlab{b}})}\BibitemShut {NoStop}%
\bibitem [{\citenamefont {Eckstein}\ \emph {et~al.}(2010)\citenamefont
  {Eckstein}, \citenamefont {Hackl}, \citenamefont {Kehrein}, \citenamefont
  {Kollar}, \citenamefont {Moeckel}, \citenamefont {Werner},\ and\
  \citenamefont {Wolf}}]{epjst2010a}%
  \BibitemOpen
  \bibfield  {author} {\bibinfo {author} {\bibfnamefont {M.}~\bibnamefont
  {Eckstein}}, \bibinfo {author} {\bibfnamefont {A.}~\bibnamefont {Hackl}},
  \bibinfo {author} {\bibfnamefont {S.}~\bibnamefont {Kehrein}}, \bibinfo
  {author} {\bibfnamefont {M.}~\bibnamefont {Kollar}}, \bibinfo {author}
  {\bibfnamefont {M.}~\bibnamefont {Moeckel}}, \bibinfo {author} {\bibfnamefont
  {P.}~\bibnamefont {Werner}}, \ and\ \bibinfo {author} {\bibfnamefont {F.~A.}\
  \bibnamefont {Wolf}},\ }\href@noop {} {\bibfield  {journal} {\bibinfo
  {journal} {Eur. Phys. J. Special Topics}\ }\textbf {\bibinfo {volume}
  {180}},\ \bibinfo {pages} {217} (\bibinfo {year} {2010})}\BibitemShut
  {NoStop}%
\bibitem [{\citenamefont {Weigert}(1992)}]{Weigert1992}%
  \BibitemOpen
  \bibfield  {author} {\bibinfo {author} {\bibfnamefont {S.}~\bibnamefont
  {Weigert}},\ }\href@noop {} {\bibfield  {journal} {\bibinfo  {journal}
  {Physica D}\ }\textbf {\bibinfo {volume} {56}},\ \bibinfo {pages} {107}
  (\bibinfo {year} {1992})}\BibitemShut {NoStop}%
\bibitem [{\citenamefont {Sutherland}(2004)}]{Sutherland2004a}%
  \BibitemOpen
  \bibfield  {author} {\bibinfo {author} {\bibfnamefont {B.}~\bibnamefont
  {Sutherland}},\ }\href@noop {} {\emph {\bibinfo {title} {{Beautiful Models:
  70 Years of Exactly Solved Quantum Many-Body Problems}}}}\ (\bibinfo
  {publisher} {World Scientific},\ \bibinfo {address} {Singapore},\ \bibinfo
  {year} {2004})\BibitemShut {NoStop}%
\bibitem [{\citenamefont {Barmettler}\ \emph {et~al.}(2009)\citenamefont
  {Barmettler}, \citenamefont {Punk}, \citenamefont {Gritsev}, \citenamefont
  {Demler}, ,\ and\ \citenamefont {Altman}}]{Barmettler2008a}%
  \BibitemOpen
  \bibfield  {author} {\bibinfo {author} {\bibfnamefont {P.}~\bibnamefont
  {Barmettler}}, \bibinfo {author} {\bibfnamefont {M.}~\bibnamefont {Punk}},
  \bibinfo {author} {\bibfnamefont {V.}~\bibnamefont {Gritsev}}, \bibinfo
  {author} {\bibfnamefont {E.}~\bibnamefont {Demler}}, , \ and\ \bibinfo
  {author} {\bibfnamefont {E.}~\bibnamefont {Altman}},\ }\href@noop {}
  {\bibfield  {journal} {\bibinfo  {journal} {Phys. Rev. Lett.}\ }\textbf
  {\bibinfo {volume} {102}},\ \bibinfo {pages} {130603} (\bibinfo {year}
  {2009})}\BibitemShut {NoStop}%
\bibitem [{\citenamefont {Lieb}, \citenamefont {Schultz},\ and\ \citenamefont
  {Mattis}(1961)}]{LSM1961}%
  \BibitemOpen
  \bibfield  {author} {\bibinfo {author} {\bibfnamefont {E.~H.}\ \bibnamefont
  {Lieb}}, \bibinfo {author} {\bibfnamefont {T.}~\bibnamefont {Schultz}}, \
  and\ \bibinfo {author} {\bibfnamefont {D.}~\bibnamefont {Mattis}},\
  }\href@noop {} {\bibfield  {journal} {\bibinfo  {journal} {Ann. Phys.
  (N.Y.)}\ }\textbf {\bibinfo {volume} {16}},\ \bibinfo {pages} {407} (\bibinfo
  {year} {1961})}\BibitemShut {NoStop}%
\bibitem [{\citenamefont {Cazalilla}\ \emph {et~al.}()\citenamefont
  {Cazalilla}, \citenamefont {Citro}, \citenamefont {Giamarchi}, \citenamefont
  {Orignac},\ and\ \citenamefont {Rigol}}]{Cazallila2011review}%
  \BibitemOpen
  \bibfield  {author} {\bibinfo {author} {\bibfnamefont {M.~A.}\ \bibnamefont
  {Cazalilla}}, \bibinfo {author} {\bibfnamefont {R.}~\bibnamefont {Citro}},
  \bibinfo {author} {\bibfnamefont {T.}~\bibnamefont {Giamarchi}}, \bibinfo
  {author} {\bibfnamefont {E.}~\bibnamefont {Orignac}}, \ and\ \bibinfo
  {author} {\bibfnamefont {M.}~\bibnamefont {Rigol}},\ }\href@noop {} {}\Eprint
  {http://arxiv.org/abs/\linebreak[0]1101.5337} {arXiv:\linebreak[0]1101.5337}
  \BibitemShut {NoStop}%
\bibitem [{\citenamefont {Haldane}(1980)}]{Haldane80}%
  \BibitemOpen
  \bibfield  {author} {\bibinfo {author} {\bibfnamefont {F.~D.~M.}\
  \bibnamefont {Haldane}},\ }\href@noop {} {\bibfield  {journal} {\bibinfo
  {journal} {Phys. Rev. Lett.}\ }\textbf {\bibinfo {volume} {45}},\ \bibinfo
  {pages} {1358} (\bibinfo {year} {1980})}\BibitemShut {NoStop}%
\bibitem [{\citenamefont {Gebhard}\ and\ \citenamefont
  {Ruckenstein}(1992)}]{Gebhard1992a}%
  \BibitemOpen
  \bibfield  {author} {\bibinfo {author} {\bibfnamefont {F.}~\bibnamefont
  {Gebhard}}\ and\ \bibinfo {author} {\bibfnamefont {A.~E.}\ \bibnamefont
  {Ruckenstein}},\ }\href {\doibase 10.1103/PhysRevLett.68.244} {\bibfield
  {journal} {\bibinfo  {journal} {Phys. Rev. Lett.}\ }\textbf {\bibinfo
  {volume} {68}},\ \bibinfo {pages} {244} (\bibinfo {year} {1992})}\BibitemShut
  {NoStop}%
\bibitem [{\citenamefont {Gebhard}\ and\ \citenamefont
  {Girndt}(1994)}]{Gebhard1994a}%
  \BibitemOpen
  \bibfield  {author} {\bibinfo {author} {\bibfnamefont {F.}~\bibnamefont
  {Gebhard}}\ and\ \bibinfo {author} {\bibfnamefont {A.}~\bibnamefont
  {Girndt}},\ }\href {\doibase 10.1007/BF01314250} {\bibfield  {journal}
  {\bibinfo  {journal} {Z. Phys. B}\ }\textbf {\bibinfo {volume} {93}},\
  \bibinfo {pages} {455} (\bibinfo {year} {1994})}\BibitemShut {NoStop}%
\bibitem [{\citenamefont {Freericks}\ and\ \citenamefont
  {Zlati{\'c}}(2003)}]{Freericks2003a}%
  \BibitemOpen
  \bibfield  {author} {\bibinfo {author} {\bibfnamefont {J.~K.}\ \bibnamefont
  {Freericks}}\ and\ \bibinfo {author} {\bibfnamefont {V.}~\bibnamefont
  {Zlati{\'c}}},\ }\href {\doibase 10.1103/RevModPhys.75.1333} {\bibfield
  {journal} {\bibinfo  {journal} {Rev. Mod. Phys.}\ }\textbf {\bibinfo {volume}
  {75}},\ \bibinfo {pages} {1333} (\bibinfo {year} {2003})}\BibitemShut
  {NoStop}%
\bibitem [{\citenamefont {Cassidy}, \citenamefont {Clark},\ and\ \citenamefont
  {Rigol}()}]{Cassidy2010a}%
  \BibitemOpen
  \bibfield  {author} {\bibinfo {author} {\bibfnamefont {A.~C.}\ \bibnamefont
  {Cassidy}}, \bibinfo {author} {\bibfnamefont {C.~W.}\ \bibnamefont {Clark}},
  \ and\ \bibinfo {author} {\bibfnamefont {M.}~\bibnamefont {Rigol}},\
  }\href@noop {} {}\Eprint
  {http://arxiv.org/abs/\linebreak[0]\linebreak[0]1008.4794}
  {arXiv:\linebreak[0]\linebreak[0]1008.4794} \BibitemShut {NoStop}%
\bibitem [{\citenamefont {Olshanii}\ and\ \citenamefont
  {Yurowsky}()}]{Olshanii2010a}%
  \BibitemOpen
  \bibfield  {author} {\bibinfo {author} {\bibfnamefont {M.}~\bibnamefont
  {Olshanii}}\ and\ \bibinfo {author} {\bibfnamefont {V.}~\bibnamefont
  {Yurowsky}},\ }\href@noop {} {}\Eprint
  {http://arxiv.org/abs/\linebreak[0]0911.5587} {arXiv:\linebreak[0]0911.5587}
  \BibitemShut {NoStop}%
\bibitem [{\citenamefont {Roux}(2010)}]{Roux2009b}%
  \BibitemOpen
  \bibfield  {author} {\bibinfo {author} {\bibfnamefont {G.}~\bibnamefont
  {Roux}},\ }\href@noop {} {\bibfield  {journal} {\bibinfo  {journal} {Phys.
  Rev. A}\ }\textbf {\bibinfo {volume} {81}},\ \bibinfo {pages} {053604}
  (\bibinfo {year} {2010})}\BibitemShut {NoStop}%
\bibitem [{\citenamefont {Danshita}\ \emph {et~al.}()\citenamefont {Danshita},
  \citenamefont {Hipolito}, \citenamefont {Oganesyan},\ and\ \citenamefont
  {Polkovnikov}}]{Danshita2010a}%
  \BibitemOpen
  \bibfield  {author} {\bibinfo {author} {\bibfnamefont {I.}~\bibnamefont
  {Danshita}}, \bibinfo {author} {\bibfnamefont {R.}~\bibnamefont {Hipolito}},
  \bibinfo {author} {\bibfnamefont {V.}~\bibnamefont {Oganesyan}}, \ and\
  \bibinfo {author} {\bibfnamefont {A.}~\bibnamefont {Polkovnikov}},\
  }\href@noop {} {}\Eprint {http://arxiv.org/abs/\linebreak[0]1012.4159}
  {arXiv:\linebreak[0]1012.4159} \BibitemShut {NoStop}%
\bibitem [{\citenamefont {Rossini}\ \emph {et~al.}(2009)\citenamefont
  {Rossini}, \citenamefont {Silva}, \citenamefont {Mussardo},\ and\
  \citenamefont {Santoro}}]{Rossini2009a}%
  \BibitemOpen
  \bibfield  {author} {\bibinfo {author} {\bibfnamefont {D.}~\bibnamefont
  {Rossini}}, \bibinfo {author} {\bibfnamefont {A.}~\bibnamefont {Silva}},
  \bibinfo {author} {\bibfnamefont {G.}~\bibnamefont {Mussardo}}, \ and\
  \bibinfo {author} {\bibfnamefont {G.~E.}\ \bibnamefont {Santoro}},\ }\href
  {\doibase 10.1103/PhysRevLett.102.127204} {\bibfield  {journal} {\bibinfo
  {journal} {Phys. Rev. Lett.}\ }\textbf {\bibinfo {volume} {102}},\ \bibinfo
  {pages} {127204} (\bibinfo {year} {2009})}\BibitemShut {NoStop}%
\bibitem [{\citenamefont {Rossini}\ \emph {et~al.}(2010)\citenamefont
  {Rossini}, \citenamefont {Suzuki}, \citenamefont {Mussardo}, \citenamefont
  {Santoro},\ and\ \citenamefont {Silva}}]{Rossini2010a}%
  \BibitemOpen
  \bibfield  {author} {\bibinfo {author} {\bibfnamefont {D.}~\bibnamefont
  {Rossini}}, \bibinfo {author} {\bibfnamefont {S.}~\bibnamefont {Suzuki}},
  \bibinfo {author} {\bibfnamefont {G.}~\bibnamefont {Mussardo}}, \bibinfo
  {author} {\bibfnamefont {G.~E.}\ \bibnamefont {Santoro}}, \ and\ \bibinfo
  {author} {\bibfnamefont {A.}~\bibnamefont {Silva}},\ }\href {\doibase
  10.1103/PhysRevB.82.144302} {\bibfield  {journal} {\bibinfo  {journal} {Phys.
  Rev. B}\ }\textbf {\bibinfo {volume} {82}},\ \bibinfo {pages} {144302}
  (\bibinfo {year} {2010})}\BibitemShut {NoStop}%
\bibitem [{\citenamefont {Harris}\ and\ \citenamefont
  {Lange}(1967)}]{Harris1967}%
  \BibitemOpen
  \bibfield  {author} {\bibinfo {author} {\bibfnamefont {A.~B.}\ \bibnamefont
  {Harris}}\ and\ \bibinfo {author} {\bibfnamefont {R.~V.}\ \bibnamefont
  {Lange}},\ }\href@noop {} {\bibfield  {journal} {\bibinfo  {journal} {Phys.
  Rev.}\ }\textbf {\bibinfo {volume} {157}},\ \bibinfo {pages} {295} (\bibinfo
  {year} {1967})}\BibitemShut {NoStop}%
\bibitem [{\citenamefont {Rela{\~n}o}(2010)}]{Relano2010a}%
  \BibitemOpen
  \bibfield  {author} {\bibinfo {author} {\bibfnamefont {A.}~\bibnamefont
  {Rela{\~n}o}},\ }\href@noop {} {}\bibinfo {howpublished} {J. Stat. Mech.
  P07016} (\bibinfo {year} {2010})\BibitemShut {NoStop}%
\bibitem [{\citenamefont {Heyl}\ and\ \citenamefont {Kehrein}()}]{Heyl2010a}%
  \BibitemOpen
  \bibfield  {author} {\bibinfo {author} {\bibfnamefont {M.}~\bibnamefont
  {Heyl}}\ and\ \bibinfo {author} {\bibfnamefont {S.}~\bibnamefont {Kehrein}},\
  }\href@noop {} {}\Eprint {http://arxiv.org/abs/\linebreak[0]1006.3522}
  {arXiv:\linebreak[0]1006.3522} \BibitemShut {NoStop}%
\end{thebibliography}

\end{document}